\documentclass{nature}

\usepackage{lineno}
\linenumbers*[1]

\usepackage{setspace}
\usepackage{amsfonts}
\usepackage{amssymb}
\usepackage{amsmath}

\usepackage{float}
\usepackage{placeins}
\usepackage{graphicx}

\usepackage[export]{adjustbox}

\usepackage[nomarkers,nolists,figuresonly]{endfloat}

\usepackage[normal,normal,bf,labelformat=empty]{caption}

\newcounter{mybodyfigure}
\newcounter{myedfigure}

\newcommand{\beginbodyfigures}{\renewcommand{\thefigure}{{\themybodyfigure}}}

\newcommand{\beginedfigures}{\renewcommand{\thefigure}{{\themyedfigure}}}

\newcommand{\stepbodyfigure}{\refstepcounter{mybodyfigure}}

\DeclareRobustCommand{\bodyfigure}[1]{\stepbodyfigure\label{#1}{\themybodyfigure}}

\newcommand{\bodyfigurelabel}[1]{\bf{Figure \bodyfigure{#1}:}}

\makeatletter
\g@addto@macro\caption@prepareslc{%
  \renewcommand{\stepbodyfigure}{\caption@l@stepcounter{mybodyfigure}}}
\makeatother

\newcommand{\stepedfigure}{\refstepcounter{myedfigure}}

\makeatletter
\g@addto@macro\caption@prepareslc{%
  \renewcommand{\stepedfigure}{\caption@l@stepcounter{myedfigure}}}
\makeatother

\usepackage[super,comma,sort&compress]{natbib}

\usepackage{hyperref}
\usepackage{verbatim}

\usepackage{ulem} 
\usepackage{rotating} 
\usepackage{soul}

\usepackage{color} 

\usepackage[colorinlistoftodos]{todonotes}

\makeatletter
\newsavebox\myboxA
\newsavebox\myboxB
\newlength\mylenA

\newcommand*\xoverline[2][0.75]{%
    \sbox{\myboxA}{$\m@th#2$}%
    \setbox\myboxB\null
    \ht\myboxB=\ht\myboxA%
    \dp\myboxB=\dp\myboxA%
    \wd\myboxB=#1\wd\myboxA
    \sbox\myboxB{$\m@th\overline{\copy\myboxB}$}
    \setlength\mylenA{\the\wd\myboxA}
    \addtolength\mylenA{-\the\wd\myboxB}%
    \ifdim\wd\myboxB<\wd\myboxA%
       \rlap{\hskip 0.5\mylenA\usebox\myboxB}{\usebox\myboxA}%
    \else
        \hskip -0.5\mylenA\rlap{\usebox\myboxA}{\hskip 0.5\mylenA\usebox\myboxB}%
    \fi}
\makeatother

\usepackage{bm}

\usepackage{comment}

\title{Alternative routes to electron hydrodynamics}

\author{Jorge Estrada-\'{A}lvarez$^{*}$, Francisco Dom\'{\i}nguez-Adame and Elena D\'{\i}az}

\let\oldequation\equation
\let\oldendequation\endequation

\renewenvironment{equation}
  {\linenomathNonumbers\oldequation}
  {\oldendequation\endlinenomath}

\let\oldsubequations\subequations
\let\oldendsubequations\endsubequations

\renewenvironment{subequations}
  {\linenomathNonumbers\oldsubequations}
  {\oldendsubequations\endlinenomath}

\begin{document}

\nolinenumbers

\maketitle

\begin{affiliations}
 
 \item GISC, Departamento de F\'{\i}sica de Materiales, Universidad Complutense, E--28040 Madrid, Spain
 
 \item[] $^{+}$To whom corresponce should be addressed: jorgestr@ucm.es

\end{affiliations}

\begin{abstract}

Viscous flow of interacting electrons in two dimensional materials features a bunch of exotic effects. A model resembling the Navier-Stokes equation for classical fluids accounts for them in the so called hydrodynamic regime. We performed a detailed analysis of the physical conditions to achieve electron hydrodynamic transport and found three new alternative routes: favouring frequent inelastic collisions, the application of a magnetic field or a high-frequency electric field. As a mayor conclusion, we show that the conventional requirement of frequent electron-electron collisions is too restrictive and, as a consequence, materials and phenomena to be described using hydrodynamics are widened. In view of our results, we discuss recent experimental evidence on viscous flow and point out alternative avenues to reduce electric dissipation in optimized devices.

\end{abstract}

\beginbodyfigures

Viscous electron flow in two-dimensional (2D) materials is a collective motion of conduction electrons~\cite{viscous_electron_fluids,hydrodynamic_approach_to_two_dimensional_electron_systems,hydrodynamic_electronic_transport,charge_transport_and_hydrodynamics_in_materials} that results in exotic phenomena such as curved current profiles\cite{visualizing_poiseuille_flow_of_hydrodynamic_electrons}, the superballistic effect~\cite{superballistic_flow_of_viscous_electron_fluid_through_graphene_constrictions} and electron whirlpools~\cite{negative_local_resistance_caused_by_viscous_electron_backflow_in_graphene,fluidity_onset_in_graphene,direct_observation_of_vortices_in_an_electron_fluid}. Both the recent realization of these counterintuitive effects and the potential applications~\cite{electronic_poiseuille_flow_in_hexagonal_boron_nitride_encapsulated_graphene_FETS,terahertz_radiation_from_the_dyakonov_shur_instability_of_hydrodynamic_electrons_in_a_corbino_geometry} in 2D devices have attracted the attention towards this field. The collective motion of electrons is fully characterized by macroscopic variables~\cite{classical_kinetic_theory_of_fluids_RESIBOIS},
by using continuum models~\cite{viscous_electron_fluids,
hydrodynamic_approach_to_two_dimensional_electron_systems,
negative_local_resistance_caused_by_viscous_electron_backflow_in_graphene,
geometric_control_of_universal_hydodynamic_flow_in_a_two_dimensional_electron_fluid,
the_boltzmann_equation_and_its_hydrodynamic_limits} similar to the Navier-Stokes equation~(NSE) for ordinary fluids, hence the name of electron hydrodynamics. The conventional route towards viscous electron flow is favouring electron-electron collisions, also known as elastic scattering events since they conserve the total momentum of the system. The requirement reads as $l_{ee} < W$ ~\cite{visualizing_poiseuille_flow_of_hydrodynamic_electrons,hydrodynamic_electronic_transport} or, more strictly with the condition $l_{ee}< l_e$ too~\cite{viscous_electron_fluids, hydrodynamic_approach_to_two_dimensional_electron_systems,negative_local_resistance_caused_by_viscous_electron_backflow_in_graphene}, where $W$ is the size of the device, $l_{ee}$ is the mean free path for elastic scattering and $l_{e}$ for inelastic scattering. The latter loses momentum after collisions with impurities and the vibrating atomic lattice. These traditional requirements restrict the materials, temperatures and systems where viscous electron flow can be reached~\cite{hydrodynamic_approach_to_two_dimensional_electron_systems}. Hence, new and less restrictive routes~\cite{hydrodynamic_electronic_transport} towards viscous flow, such as the recently discovered para-hydrodynamics~\cite{direct_observation_of_vortices_in_an_electron_fluid,parahydrodynamics_from_weak_surface_scattering_in_ultraclean_thin_flakes}, are desirable since they would facilitate the development of new applications such as less energy-demanding devices.


The aim of this work is to rigorously explore the requirements for collective behaviour of the electrons. By solving the Boltzmann transport equation (BTE), which is assumed to be exact as described in Section \ref{BTEvalidity} of the supplementary information (SI), the electron distribution is obtained. Transport may be collective or not, depending on the particular magnitude of the length scales $W$, $l_e$, $l_{ee}$, and eventually, the cyclotron radius $l_B$
 due to a magnetic field and a length scale $l_\omega$ associated to an ac driving, as well as on the edge scattering properties. 
 
 When the electronic behaviour is collective~\cite{classical_kinetic_theory_of_fluids_RESIBOIS,equilibrium_and_nonequilibrium_statistical_mechanics_BALESCU,kinetic_theory_and_transport_phenomena_SOTO}, the NSE 
 gives correct predictions for macroscopic variables such as the drift velocity. 
 It is worth mentioning that 
 although the formal derivation of the NSE in conventional fluids is based on the conservation of the number of particles, the momentum and the energy of the
fluid, in solid-state systems, the unavoidable electron interaction with phonons and defects implies that the total momentum is not conserved. The latter is usually considered by including a new dissipative term
in a modified condensed-matter NSE~\cite{viscous_electron_fluids,hydrodynamic_approach_to_two_dimensional_electron_systems,negative_local_resistance_caused_by_viscous_electron_backflow_in_graphene,geometric_control_of_universal_hydodynamic_flow_in_a_two_dimensional_electron_fluid}. As a consequence such term extends the validity of this modified NSE to account for the diffusive regime of transport, that cannot be considered as hydrodynamics. In this regime the considered NSE and the well-known Drude model (DRE) give rise to the same results. In our work we propose the accuracy of the NSE and the DRE, which are assessed by comparison with the BTE results, as a fundamental and quantitative criterion for viscous electron flow. Collective behaviour is associated with accurate NSE predictions but unaccurate DRE predictions, as otherwise it would be diffusive. Moreover, it is robust regardless of choosing the total current, the velocity or the Hall profile as the observable of interest. In this paper, we study the requirements for viscous electron flow in terms of the characteristic physical length scales and we discuss how the alternative routes affect the most remarkable hydrodynamic signatures above mentioned.

\section*{Theoretical models}

\subsection{Boltzmann transport equation}

Consider a 2D system where electrons behave as semiclassical particles~\cite{ashcroft_solid_state_physics,di_ventra_electrical_transport_in_nanoscale_systems}, with a well-defined position vector ${\bm r}=(x,y)$ and wave vector ${\bm k}=(k_x,k_y)$, and let $\hat{f}({\bm r},{\bm k},t)$ be their distribution function that obeys the BTE~\cite{di_ventra_electrical_transport_in_nanoscale_systems,ballistic_and_hydrodynamic_magnetotransport_in_narrow_channels, ballistic_to_hydrodynamic_transition_and_collective_modes_for_two_dimensional_electron,higher_than_ballistic_conduction_of_viscous_electron_flows}
\begin{equation}
\partial_t \hat{f}  + {\bm v} \cdot \nabla_{{\bm r}}  \hat{f} - \frac{e}{\hbar} \, \big( -\nabla  \hat{V} + {\bm v} \times {\bm B} \big) \cdot \nabla_{{\bm k}}  \hat{f} =  \Gamma \left[  \hat{f} \right]\ ,
\label{BTEtimeDependent}
\end{equation}
where ${\bm v} = \hbar {\bm k} / m$ and $-e$ are the electron's velocity and charge, respectively, $\hbar$ the reduced Plank constant and $m$ the effective mass. Electrons experience a Lorentz force due to a time-harmonic electric potential $\hat{V}({\bm r},t) = V({\bm r}) e^{i \omega t}$, either set at the contacts or with an electromagnetic wave of frequency $\omega$, and a perpendicular magnetic field ${\bm B} = B \hat{\bm z}$. $\Gamma [ \hat{f} ]$ is the collision operator, including all sources of electron scattering. Under Callaway's ansatz~\cite{Callaway1959,probing_carrier_interactions_using_electron_hydrodynamics,hyrdodynamic_electron_flow_in_high_mobility_wires}, the collision term splits as
$\Gamma [ \hat{f} ] = - ({\hat{f} - \hat{f}^e})/{\tau_e} -  ({\hat{f} - \hat{f}^{ee} })/{\tau_{ee}}$, 
where $\tau_e$ is the relaxation time for inelastic collisions with impurities and phonons towards the equilibrium distribution $\hat{f}^e$, and $\tau_{ee}$ accounts for elastic collisions with other electrons towards the local distribution $ \hat{f}^{ee}$ shifted by the electron drift velocity. The latter may also describe the effect of the para-hydrodynamics reported in Ref.~\cite{parahydrodynamics_from_weak_surface_scattering_in_ultraclean_thin_flakes}. The intermediate tomographic regime~\cite{tomographic_dynamics_and_scale_sependent_viscosity_in_2D_electron_systems,collective_modes_in_interacting_two_dimensional_tomographic_fermi_liquids} does not obey Callaway's ansatz, but it may described assuming two different values for the relaxation rate $\tau_{ee}$ (see Section~\ref{tomographicSection} of the SI).

We assume the electron density $n$ to be constant and set by a gate potential, so that $\hat{f}^e$ does not depend on ${\bm r}$ or $t$. Let $k_F$ be the Fermi wavenumber and ${\rm v}_F = \hbar k_F/m$ the Fermi velocity in an isotropic band structure. We rewrite ${\bm k} = k \hat{\bm u}_{\bm k}$, with $\hat{\bm u}_{\bm k}\equiv \left (\cos \theta , \sin \theta \right)$, and define $\hat{g}({\bm r},\theta,t) \equiv (4\pi^2 \hbar/m)\!\! \int_{0}^\infty (\hat{f} - \hat{f}^e)\, {\rm d} k$ and $\hat{g}^{ee}({\bm r},\theta,t) \!\equiv\! (4\pi^2 \hbar/ m)\!\!\int_{0}^\infty(\hat{f}^{ee} - \hat{f}^e ) {\rm d} k$.
Moreover $\int_{0}^{2\pi} \hat{g}({\bm r},\theta,t) {\rm d} \theta = 0$ and we assume $|\hat{g} | \ll {\rm v}_F$, namely, phenomena happen near the Fermi surface. We integrate Eq. ~\eqref{BTEtimeDependent} over $k$ and look for a solution $\hat{g}({\bm r},\theta,t)=\Re [ g({\bm r},\theta ) e^{i \omega t}] $, where $\Re$ stands for the real part. The following equation holds for the length scales defined in Table~\ref{tab:lengthScales}
\begin{equation}
i\,\frac{g}{l_\omega} + \hat{\bm u}_{\bm k}
\cdot \nabla_{{\bm r}} \left( g- \frac{e V}{m {\rm v}_F} \right) + \frac{\partial_\theta g}{l_B} + \frac{g}{l_e} + \frac{g-g^{ee}}{l_{ee}} = 0 \ ,
\label{BTE}
\end{equation}  
where $g^{ee} \simeq u_x \cos \theta + u_y \sin \theta $ and the components of the drift velocity are $u_x({\bm r}) = (1/\pi) \int_0^{2\pi} g({\bm r},\theta) \cos \theta \, {\rm d} \theta$ and $u_y({\bm r}) = (1/\pi) \int_0^{2\pi} g({\bm r},\theta) \sin \theta \, {\rm d} \theta$.

Previous works have proposed a formalism based on non-local conductivity tensors to solve the BTE and particularly to analyze the hydrodynamic regime~\cite{parahydrodynamics_from_weak_surface_scattering_in_ultraclean_thin_flakes,robustness_of_vorticity_in_electron_fluids}. However, although formally such approach can be used in arbitrary geometries, its practical application has been reduced so far to limited edges conditions. In this regard, our current proposal overpass this limitation.



\subsection{The Navier-Stokes equation}

Let us ignore higher modes of $g(\theta)$ to find a model equation based on macroscopic variables, this is, the drift velocity $(u_x,u_y)$ and $(w_x, w_y)$ which is related to the stress tensor in classical hydrodynamics. We write $g$ as a distribution depending on these variables~\cite{ballistic_and_hydrodynamic_magnetotransport_in_narrow_channels}
\begin{equation}
g = u_x \cos  \theta + u_y \sin \theta + w_x \cos 2 \theta + w_y \sin 2 \theta\ .
\label{armonics2}
\end{equation}
By considering this level of approximation and following the procedure detailed in Section~\ref{NSEfromBTE} of the SI, Eq.~\eqref{BTE} can be recast as
\begin{subequations}
\begin{eqnarray}
\nabla \cdot {\bm u} & = & 0 \ , \label{NScontinuity}\\
- \nu \nabla^2 {\bm u} + \left( \omega_B +\nu_H \nabla^2 \right) {\bm u} \times {\hat{\bm z}} + \left(\frac{\mathrm{v}
_F}{l_e} + i \omega\right){\bm u}& = &
\dfrac{e}{m}\,\nabla V\ ,    \label{NS_momentum} 
\end{eqnarray}
\label{NS}%
\end{subequations}
%
%
that resemble the continuity equation and the NSE for classical fluids~\cite{viscous_electron_fluids,hydrodynamic_approach_to_two_dimensional_electron_systems,negative_local_resistance_caused_by_viscous_electron_backflow_in_graphene,geometric_control_of_universal_hydodynamic_flow_in_a_two_dimensional_electron_fluid}. 
We define the cyclotron frequency $\omega_B \equiv e B/m$, the viscosity $\nu$ and Hall viscosity $\nu_H$ as follows~\cite{negative_magnetoresistance_in_viscous_flow_of_two_dimensional_electrons,measuring_hall_viscosity_of_graphene_electron_fluid}
\begin{equation}
\nu \equiv  \frac{{\rm v}_F \left( l_e^{-1} + l_{ee}^{-1} + i l_\omega^{-1} \right) }{4 \left( l_e^{-1} + l_{ee}^{-1} + i l_\omega^{-1} \right)^{2} + 16 l_B^{-2}} \ ,  \hspace{1cm} \nu_H \equiv \frac{{\rm v}_F {l}_{B}^{-1} }{2 \left( l_e^{-1} + l_{ee}^{-1} + i l_\omega^{-1} \right)^{2} + 8 l_B^{-2}}\ .
\label{viscosity}
\end{equation}
We notice that these definitions take into account the contribution of all considered effects, such as those derived from inelastic collisions. 
Including $l_e$ in this expression is not new~\cite{hyrdodynamic_electron_flow_in_high_mobility_wires} and it is similar to the effects considered by the ballistic correction proposed in previous works~\cite{superballistic_flow_of_viscous_electron_fluid_through_graphene_constrictions,higher_than_ballistic_conduction_of_viscous_electron_flows,geometric_control_of_universal_hydodynamic_flow_in_a_two_dimensional_electron_fluid}. However in the current proposal it is not a phenomenological correction but remarkably it arises naturally from our model. Notice that Eq.~\eqref{viscosity} has a dissipative term in $\bm u$ arising from non-conserving-momentum collisions ($l_e<\infty$), and this is why it contains the Drude equation~(DRE) as a particular case.

\subsection{Sample boundaries.}

Boundary conditions that account for edge scattering are crucial to properly describe viscous electron flow within any model~\cite{boundary_conditions_of_viscous_electron_flow}. It is generally admitted that a direct comparison between the NSE and the BTE approaches is clearly not trivial. The main difficulty is how to deal with boundary conditions in such a way that both methods could treat the same transport regimes. In the present work, one of the most relevant and innovative achievements is the development of a well-defined formalism to deal with the same edge scattering properties within both models, see Section~\ref{BoundaryCondition} of the SI. Since the particular boundary details strongly depend on the experimental conditions, we shall study the most currently accepted types of boundary conditions derived from microscopical considerations. First, we consider a fully diffusive (DF) edge where incident electrons are scattered in all directions regardless of their angle of incidence \cite{visualizing_poiseuille_flow_of_hydrodynamic_electrons}. Second, we analyze a partially specular (PS) edge where scattering is due to the irregularities of the boundary. Notice that this scenario is affected by the boundary roughness defined by way of its dispersion coefficient $d \equiv \sqrt{\pi} h^2 h'k_F^3 \lesssim 1$, where $h$ is the edge's bumps mean height and $h'$ is its correlation length. The exact expressions for the BTE and the derivation of the NSE conditions in terms of the so-called slip length $\xi$~\cite{boundary_conditions_of_viscous_electron_flow}
are reported in Section~\ref{BoundaryCondition} of the SI and result in the following definitions
\begin{equation}
    \xi = \begin{cases} \dfrac{3\pi }{4}\dfrac{\nu}{{\rm v}_F}\ , & \text{DF edge}\ , \\ 
     8\left( \dfrac{1}{d} - \dfrac{2}{3\pi } \right)\dfrac{\nu}{{\rm v}_F}\ , & \text{PS edge} \ . \end{cases}
    \label{slipLength}
\end{equation}
Notice that the validity condition $d\lesssim 1$ prevents the occurrence of negative values of the slip length. These results are compatible with previous studies where the effect of inelastic collisions, magnetic and ac electric fields were neglected~\cite{boundary_conditions_of_viscous_electron_flow}. Here, a flat PS edge ($d=0$) leads to the perfect slip boundary condition $\xi \to \infty $, while the no-slip condition $\xi = 0$ remains beyond the microscopic approach. 

\section*{Results}


\begin{figure}[t]
\centerline{\includegraphics[width=0.65\columnwidth]{ 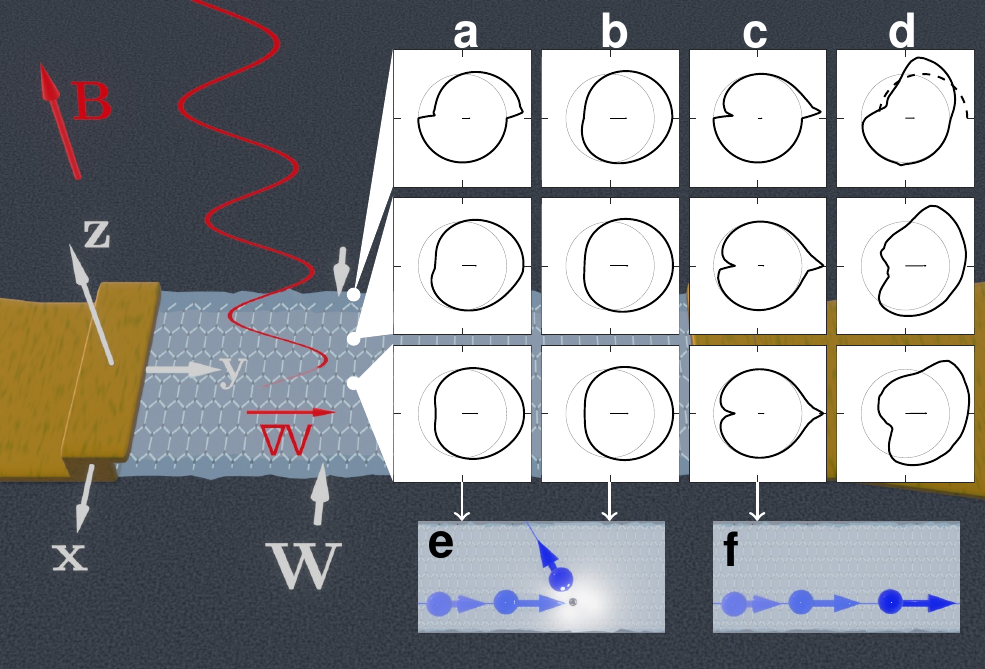}}
\caption{\bodyfigurelabel{fig:pseudoFermi} Sketch of a long graphene ribbon of width $W$ where electrons flow under the effect of a potential gradient $\nabla V$ and a magnetic field ${\bm B} = B \hat{\bm z}$. Panels (a)-(d) show polar plots of the distribution $g(\theta)$ derived from the BTE for three different positions $x=0, -W/4$ and $-W/2$. Deviations between $g(\theta)$ and  the equilibrium distribution (grey circle) has been enlarged for clarity. (a) Intermediate regime where $l_e = l_{ee} = W$, $l_B, l_\omega \to \infty$ and a DF edge. (b)~Same parameters as in panel~(a) but for a PS edge with $d = 0.5$. (c)~Ballistic regime where $l_e = l_{ee} = 10 W$ and $l_B, l_\omega \to \infty$. (d)~Commensurability effect where $l_e = l_{ee} = 10W$, $l_B = W$, $l_\omega \to \infty$ and a DF edge. Notice the lack of incident electrons in some directions in comparison with the dashed line (top panel). Panels~(e) and ~(f) show a schematic representation of the electron velocity for the situations considered in panels~(a), (b) and (c) respectively. Electron scattering in~(e) makes the NSE works properly. Ballistic electrons move parallel to the channel with increasing momentum, as depicted in panel~(f). Section~\ref{PolarExtended} of the SI presents polar plots of the distribution $g(\theta)$ for an enlarged set of parameters.}
\end{figure}


Figure~\ref{fig:pseudoFermi} shows the distribution $g(\theta )$ given by the BTE for three different positions inside a very long channel under a constant potential gradient $\partial_y V$. Let us start discussing the role of the boundary conditions in an intermediate regime transport such that $l_e = l_{ee} = W$, $l_B$ and $l_\omega \to \infty$. Figure~\ref{fig:pseudoFermi}(a) shows the distributions for a DF edge~[see Eq.~\eqref{boundaryBTEdiffusive} in the SI] where electrons are uniformly scattered in all directions, while Fig.~\ref{fig:pseudoFermi}(b) corresponds to a PS edge~[see Eq.~\eqref{boundaryBTEspecular} in the SI]. The angular distribution at the edge is less sharp when there is some specularity. Hence, regarding the truncated harmonic expansion~\eqref{armonics2}, we expect the NSE to be more accurate for PS edges than for DF edges. 
Concerning the ballistic regime ($l_e = l_{ee} = 10 W$), Figs. ~\ref{fig:pseudoFermi}(c) and (d) present $g(\theta )$ in absence ($l_B \to \infty$) and in the presence of a commensurable magnetic field ($l_B = W$), respectively. Although some electrons collide against the walls in the first case, others travel almost parallel to the channel without collisions and continuously increase their momentum. Such an accumulation of electrons traveling parallel to the channel, depicted in Fig.~\ref{fig:pseudoFermi}(f), accounts for the sharp peak that appears in the pseudo-Fermi distribution in Fig.~\ref{fig:pseudoFermi}(c). The NSE level of approximation based on the consideration of the first two angular harmonics Eq.~\eqref{armonics2} is clearly not compatible with such $g(\theta)$. Conversely, elastic and inelastic collisions keep these electrons from gaining momentum, as depicted in Fig.~\ref{fig:pseudoFermi}(e). Moreover, they relax the angular distribution to smoother functions, as presented in Fig.s~\ref{fig:pseudoFermi}(a) and~\ref{fig:pseudoFermi}(b), and make the NSE accurate when $l_e < W$ or $l_{ee} < W$. 
Interestingly, this is not the case in the so-called tomographic regime~\cite{tomographic_dynamics_and_scale_sependent_viscosity_in_2D_electron_systems,collective_modes_in_interacting_two_dimensional_tomographic_fermi_liquids}, as shown in Section~\ref{tomographicSection} of the SI, and thus, it cannot be considered as corresponding to a viscous electron flow.

\begin{figure}[h]
\centerline{\includegraphics[width=0.65\columnwidth]{ 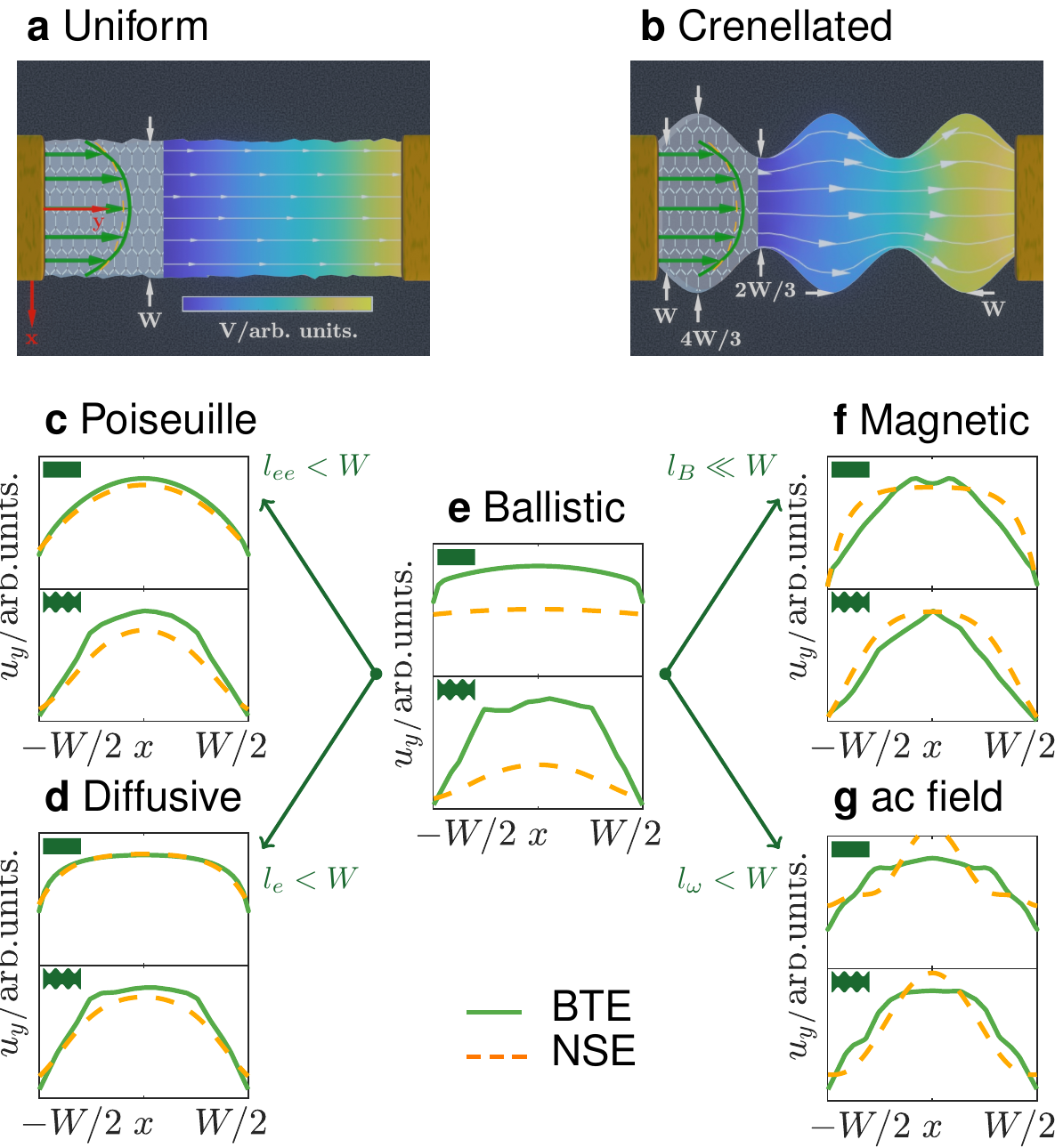}}
\caption{\bodyfigurelabel{fig:profiles_velocity}  Electron flow simulations in a very long uniform channel and a crenellated one with DF boundaries. 
Panels (a) and (b) show the BTE potentials and electron streamlines in the Poiseuille regime. (c)~Profiles of the drift velocity along the channel with frequent elastic collisions ($l_{ee} = 0.25 W$) is the usual route to viscous electron flow and result in a Poiseuille flow. 
(d)~Profiles  with frequent inelastic collisions ($l_e = 0.25 W$).
(e)~Ballistic regime where the NSE results in wrong predictions. Here $l_e = l_{ee} = 10 W$ for the uniform channel and $l_e = l_{ee} = 2 W$ for the crenellated one. In the latter the uneven walls have an stronger effect 
on the electron flow and thus it presents ballistic behaviour even for slightly shorter $l_e$ and $l_{ee}$. 
(f)~Profiles in the presence of a magnetic field ($l_B = 0.25 W$).
(g)~Profiles in the presence of an ac field ($l_\omega = 0.25 W$). This Fig. shows the accuracy of the NSE to reproduce the curved velocity profile. Section~\ref{ProfileExtended} of the SI presents streamlines and velocity profiles for an enlarged set of parameters.}
\end{figure} 

As shown in previous works~\cite{visualizing_poiseuille_flow_of_hydrodynamic_electrons,ballistic_and_hydrodynamic_magnetotransport_in_narrow_channels}, the curvature of the current density profile in a uniform channel might not be a good indicator for hydrodynamic behaviour. However, our approach can be tested in such a physical scenario, as well as in other non-uniform geometries, to evaluate the deviation between the results from the NSE and the BTE instead. Figure \ref{fig:profiles_velocity} shows the velocity profiles in uniform~(a) and crenellated~(b) channels for different values of the length scales of the system. The main conclusion of Fig. ~\ref{fig:profiles_velocity} is that the NSE and the BTE predictions are completely different in the ballistic regime, but they almost overlap in the Poiseuille, diffusive, magnetic and ac field panels. 

Additionally our results allows us to also understand how all the considered effects result in curved profiles.
A Poiseuille profile, which is almost parabolic, emerges when $l_{ee} \ll W \ll l_e$, as depicted in Fig.~\ref{fig:profiles_velocity}(c). Nevertheless, a nonzero curvature of the velocity profile is not unique to this regime. Even in the diffusive regime $l_e \ll W$ shown in Fig. \ref{fig:profiles_velocity}(d), and in the absence of elastic collisions $l_{ee} \gg W$, the profile is not flat.
The hydrodynamic behaviour and curvature of the transverse velocity profile is proven in presence of a magnetic field and an ac field in Fig.s~\ref{fig:profiles_velocity}(f) and (g) respectively. Although previously suggested~\cite{ballistic_hydrodynamic_phase_transition_in_flow_of_two_dimensional_electrons}, we prove that the former is a stronger hydrodynamic fingerprint by a direct comparison between the BTE and the NSE. Most remarkably, the fact that hydrodynamic features can also be induced by the application of an ac field is a new promising result. For completeness Section~\ref{ProfileExtended} of the SI shows the prediction of the velocity profile in an extended set of transport regimes, where in addition the DRE solution is shown as a wrong approximation to the problem. A detailed derivation of the~DRE is presented in Section~\ref{DRE_validity} of the SI.


\begin{figure*}
\centerline{\includegraphics[width=0.95\columnwidth]{ 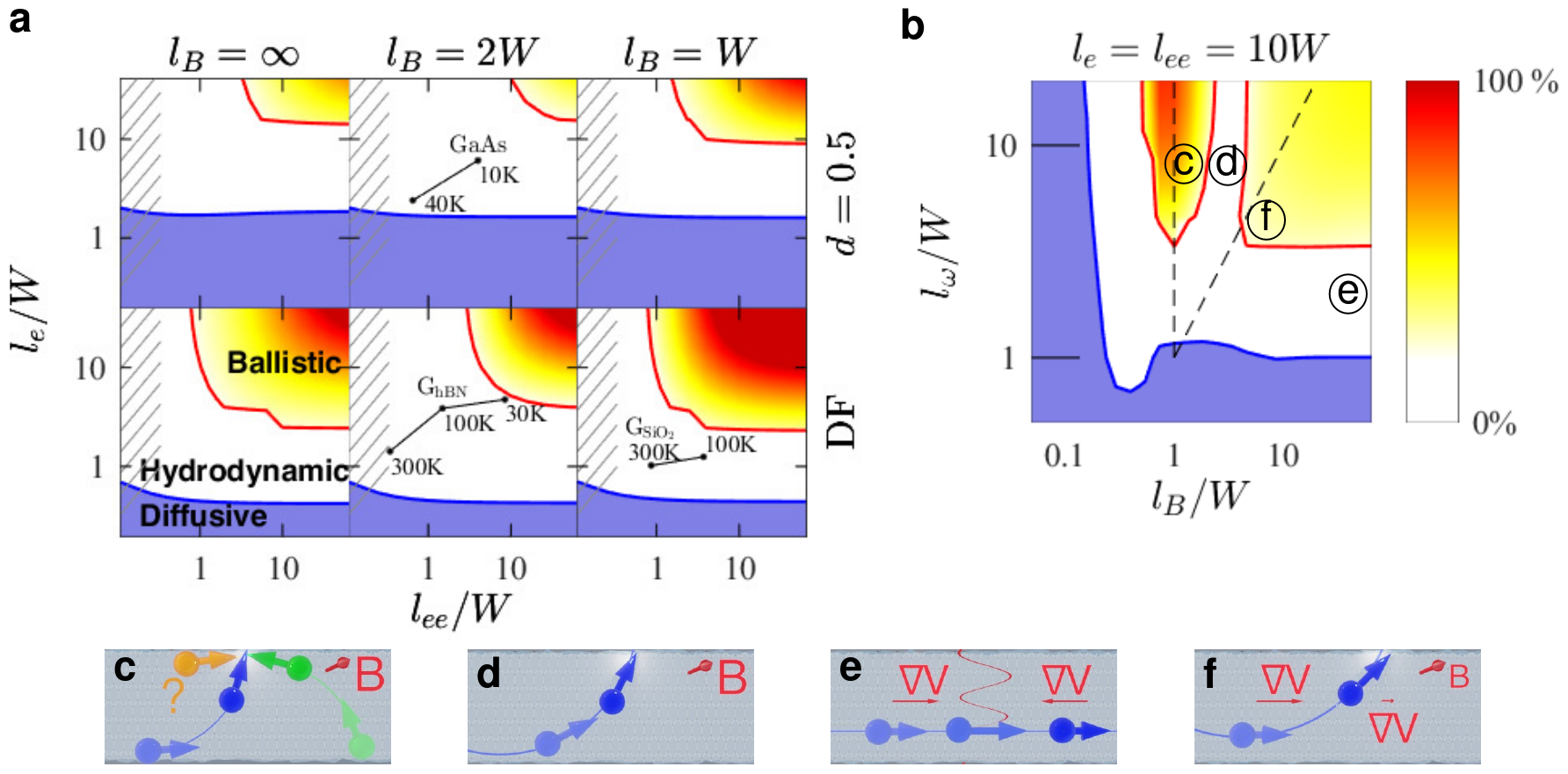}}
\caption{\bodyfigurelabel{fig:errorPlots} Panels (a) and (b) show color maps that represent the model error of the total current obtained with the NSE in comparison with the BTE results. White regions remark the hydrodynamic regime where only NSE results are correct and blue regions are those where both the NSE and the DRE account for the diffusive regime of transport. Red (blue) contour lines limit the region where de NSE (DRE) error is 20\%. (a) NSE error versus $l_e$ and $l_{ee}$ mean free paths, in absence of an ac field. Each plot accounts for a DF or PS ($d = 0.5$) edge, and for a different cyclotron radius. Grey patterned areas, defined by the conditions $l_{ee} < W/3$ and $l_{ee} < l_e / 3$, indicate the conventional requirements ($l_{ee} \ll W$ and $l_{ee} \ll l_e$) to achieve the hydrodynamic regime. For comparison, overprinted lines show typical physical conditions considered in some experiments in high quality graphene (G$\mathrm{_{hBN}}$) in a $W = 500 \rm \, nm$-wide channel~\cite{superballistic_flow_of_viscous_electron_fluid_through_graphene_constrictions,the_electronic_properties_of_graphene}, graphene on a $\mathrm{SiO}_2$ substrate (G$\mathrm{_{SiO_2}}$)~\cite{boron_nitride_substrates_for_high_quality_graphene_electronics} and GaAs (GaAs)~\cite{geometric_control_of_universal_hydodynamic_flow_in_a_two_dimensional_electron_fluid} in a $W = 200 \rm \, nm$ channel (see Section \ref{ModelParameters} of the SI). (b)~NSE error versus the cyclotron radius $l_B$ and the ac length $l_\omega$ for $l_e = l_{ee} = 10W$ and a DF edge. Commensurability resistance happens when $l_B \simeq W$ and cyclotron resonance when $l_B \simeq l_\omega$. Panels~(c)--(f) represent electrons as semiclassical particles moving under a potential gradient $\nabla V$ and a perpendicular magnetic field $B$, for some relevant positions labeled in panel~(b). (c)~Commensurability effect arises when electrons cannot reach the boundaries in all directions. (d) Electrons do not follow straight trajectories under a magnetic field. (e)~An electron moving parallel to the channel and subject to an ac electric field will alternately accelerate and decelerate due to the field swiping direction. (f)~The magnetic field and the ac electric field mutually hinder when $l_\omega=l_B$, namely the resonant condition $\omega=\omega_B$ is met.
Section~\ref{ErrorExtended} of the SI shows error maps evaluated by means of the total electriccurrent for an enlarged set of parameters.}
\end{figure*}

Let us now focus on the deviations of the NSE predictions from the BTE results. To this end, we calculate the electric current $I_\mathrm{NSE}$ and $I_\mathrm{BTE}$ from the velocity profile obtained after solving the NSE and the BTE, respectively, and define the relative error as $\epsilon=2|I_\mathrm{NSE}-I_\mathrm{BTE}|/|I_\mathrm{NSE}+I_\mathrm{BTE}|$, as explained in Methods. The electric current is proportional to the area under each velocity profile, and it is a relevant quantity in most experiments~\cite{visualizing_poiseuille_flow_of_hydrodynamic_electrons,imaging_hydrodynamic_electrons_flowing_without_landauer_sharvin_resistance,simultaneous_voltage_and_current_density_imaging_of_flowing_electrons_in_two_dimensions}. The accuracy of the DRE has been accounted accordingly in Section~\ref{ErrorExtended} of the SI to monitor the parameters that lead to identical results of the BTE, NSE and DRE, namely, the diffusive regime of transport.

Figure~\ref{fig:errorPlots}(a) shows the NSE error when compared to the BTE in the presence of a magnetic field and considering different boundary conditions. Notice that the grey patterned area corresponds to the conventional requirements for hydrodynamic onset. The red contour line limits the region where the NSE error is larger than $20\%$, defined as the ballistic regime of transport. Similarly the blue contour line surrounds the diffusive regime (blue area) such as the DRE error is smaller than $20\%$. We remark that these results would be similar if we choose a threshold within $10\%$ and $30\%$ to define the different transport regimes. Thus, white regions account for reliable sets of parameters supporting hydrodynamic transport. In all considered cases, our results demonstrate the validity of the NSE beyond the usually accepted conditions (see white areas). Indeed, the increase of the magnetic field also benefits the hydrodynamic regime as demonstrated in Fig.~\ref{fig:profiles_velocity}(a). Still, the new region of validity slightly shrinks for the particular case $l_B \simeq W$. The latter occurs due to the already known commensurability effect of resistance~\cite{boundary_scattering_in_ballistic_graphene}, that some works referred to as a phase transition~\cite{ballistic_hydrodynamic_phase_transition_in_flow_of_two_dimensional_electrons}. In this particular situation, cyclotron orbits prevent electrons coming from a particular edge to reach the opposite one with all angles, as shown in Fig.~\ref{fig:errorPlots}(c). Furthermore, Fig.~\ref{fig:pseudoFermi}(d) proves that this effect depletes entire regions of the pseudo-Fermi distribution, and Eq.~\eqref{ansatzWall} for the incident distribution at the wall fails. Therefore, as a general trend, increasing the magnetic field while keeping $l_B \nsim W$  makes the NSE valid, as shown in Fig.~\ref{fig:errorPlots}(a). Still there is a field threshold above which the system will eventually enter the diffusive regime of transport. Indeed, a Lorentz force, not balanced by a Hall field, prevents electrons from following straight trajectories, as illustrated in Fig.~\ref{fig:errorPlots}(d).

We also study the effects of an ac electric field polarized along the channel when a magnetic field is also applied. Then there is an overall decrease in the NSE's error when $l_e \gg W$ and $l_{ee} \gg W$, shown in Fig.~\ref{fig:errorPlots}(b). If we consider electrons traveling parallel to the channel, while they are increasing momentum the ac field swaps its direction leading to slow them down, as shown in Fig.~\ref{fig:errorPlots}(e). Thus, inertia avoids the formation of peaks such as those highlighted in Fig.~\ref{fig:pseudoFermi}(c) and makes the NSE valid under an ac field, as also demonstrated in Fig.~\ref{fig:errorPlots}(b). However, we find an increase in the error due to the cyclotron resonance condition $l_B \simeq l_\omega$ ($\omega_B \simeq \omega$), so that both effects couple and mutually hinder, as depicted in Fig.~\ref{fig:errorPlots}(f).  Fig.~\ref{fig:errorPlots} also sheds light on the transition from the diffusive to the ballistic regime when changing the various length scales involved in charge transport. It is apparent from Fig.~\ref{fig:errorPlots} that, in almost all cases studied, the transition is not abrupt but an intermediate transport regime is reached, where the NSE describes accurately non-equilibrium collective electron dynamics (see white areas).

It is worth noticing that some authors suggest the Hall field instead of the current density profile to assess the hydrodynamic regime~\cite{visualizing_poiseuille_flow_of_hydrodynamic_electrons,ballistic_and_hydrodynamic_magnetotransport_in_narrow_channels}. It is indeed remarkable that from our novel perspective (comparison between NSE and BTE), the consideration of the current profiles, the Hall voltage, the total current or the total Hall voltage yield the same conclusions, as proven in Section~\ref{HallVoltage} of SI.

In summary, according to the analysis of the transverse velocity profiles (see Fig.~\ref{fig:profiles_velocity}) and the evaluated error of the NSE in comparison with the BTE (see Fig.~\ref{fig:errorPlots}), we demonstrate that the traditional requirement for hydrodynamic transport in solid-state systems, $l_{ee}<W$, can be overcome by the the new alternative routes whose general trends are summarized in Fig.~\ref{fig:errorPlots}: i) keep the inelastic scattering length within the order of the device size, $l_e\sim W$, ii) include a magnetic field such that a $l_B$ is the same order of magnitude as $W$ but different from the commensurability condition and iii) the application of an ac field such that a $l_\omega \sim W$.

Error maps shown in Fig.~\ref{fig:errorPlots}(a) also include typical physical conditions considered in some experiments for comparison~\cite{geometric_control_of_universal_hydodynamic_flow_in_a_two_dimensional_electron_fluid,superballistic_flow_of_viscous_electron_fluid_through_graphene_constrictions,
the_electronic_properties_of_graphene}. Section~\ref{ModelParameters} of the SI reports on the detailed description of the considered sets of parameters. Notice that although some materials, such as low quality graphene G$\mathrm{_{SiO_2}}$, never enter the conventionally accepted region, G$\mathrm{_{SiO_2}}$ narrow channels may enter the hydrodynamic regime by way of the new alternative routes. Our analysis changes the widely accepted paradigm and proves that more materials and more temperatures can be studied using hydrodynamics tools. This also establishes that the ratio $l_{ee}/l_e$ is not so crucial concerning the validity of the NSE. As a reference, let us remark that for a graphene ribbon of width $W = 500 \, \rm nm$ at $n = 10^{12} \, \rm electrons/cm^2$, $l_B = 500 \, \rm nm$ ($B \simeq 250 \, \rm mT$), while $l_\omega = 500 \, \rm nm$ ($\omega / 2\pi \simeq 0.3 \, \rm THz$), so both $l_B \sim W$ and $l_\omega \sim W$, are plausible requirements. 

%

\section*{Discussion}

The new hydrodynamic routes proposed in this work shed light on some of generally admitted signatures of viscous electron flow. Let us focus first on non-local experiments based on uniform Hall bars in the so-called proximity geometry~\cite{negative_local_resistance_caused_by_viscous_electron_backflow_in_graphene,superballistic_flow_of_viscous_electron_fluid_through_graphene_constrictions,fluidity_onset_in_graphene}. In Ref.~[\citenum{fluidity_onset_in_graphene}], the authors demonstrate the existence of a transition from ballistic to hydrodynamic transport regimes based on the occurrence of a sharp maximum in the negative resistance. Notice that this effect is also related to the existence of unexpected backflow or even small whirlpools due to the viscous nature of the electron flow. The conventional hydrodynamic description, which requests the condition $l_{ee} < W$, agrees with their measurements at intermediate temperatures. However, the fact that the hydrodynamic onset survives at very low temperatures when $l_{ee}\to \infty$, is incompatible with $l_{ee} < W$. Remarkably, the latter is indeed compatible  with our results, since we have demonstrated that the condition $l_e \sim W$ is already sufficient for the hydrodynamic transport to occur (see Section~\ref{Geometry} of the SI). The fact that the hydrodynamic regime only exists at low carrier density is consistent with $l_e$ decreasing near the neutrality point~\cite{negative_local_resistance_caused_by_viscous_electron_backflow_in_graphene}. 

The alternative route $l_e \sim W$ is also supported by the  direct visualization of the Poiseuille flow of an electron fluid~\cite{visualizing_poiseuille_flow_of_hydrodynamic_electrons}. Here, the Hall field profile across a high-mobility graphene channel is used as the key for distinguishing ballistic from hydrodynamic flow. A curved profile is not unique to the conventional situation $l_{ee} < W$, as it also arises when $l_e \sim W$, in agreement with Fig.~\ref{fig:profiles_velocity}(c). Within the same experimental setup, another alternative route with the magnetic field is made clear. Indeed, the authors find a sharp increase in the profile curvature when increasing the magnetic field,
showing that our proposed condition to tune $l_B$ triggers the hydrodynamic onset of the NSE. 

Last let us comment on the experimental hydrodynamic evidence known as superballistic conduction. This regime of collective transport refers to devices with a resistance under its ballistic limit~\cite{superballistic_flow_of_viscous_electron_fluid_through_graphene_constrictions,higher_than_ballistic_conduction_of_viscous_electron_flows,geometric_control_of_universal_hydodynamic_flow_in_a_two_dimensional_electron_fluid}. We notice the superballistic regime cannot be exclusively related to the case of frequent elastic collisions, since it can also be reached with the proposed alternative routes as shown in  Fig.~\ref{fig:superballistic}. Particularly under the condition $l_B \ll W$, such phenomenon is known as negative magnetoresistance~\cite{negative_magnetoresistance_in_viscous_flow_of_two_dimensional_electrons,geometric_control_of_universal_hydodynamic_flow_in_a_two_dimensional_electron_fluid}.
Other experiments~\cite{boundary_scattering_in_ballistic_graphene} show how, after the resistance peak due to the commensurability effect ($l_B = W$), a further increase of the magnetic field $l_B \ll W$ results in a resistance under the ballistic limit. 

\section*{Conclusion}

To conclude, in this work we develop a framework to approximate the general BTE to the simplified NSE and to define where electron transport is hydrodynamic. We believe that our approach has several advantages, mainly because it is rigorously based on the requirements for collective behaviour as a fundamental premise. This allows us to perform our analysis with no need of initial assumptions for the viscosity, so that its dependence on all length scales $(l_e, l_{ee}, l_B, l_\omega,W)$ arises naturally. In addition, our approach is not directly related to geometrical constraints so it can be applied in many other physical scenarios. 
We conclude that the widely admitted requirements for viscous electron flow are too restrictive and limit hydrodynamic transport to very particular experimental conditions, regarding temperature range or materials quality. Remarkably, the new alternative routes actually lead to the most noteworthy hydrodynamic signatures. Our novel proposal to search for hydrodynamic signatures for ac electric fields~\cite{terahertz_radiation_from_the_dyakonov_shur_instability_of_hydrodynamic_electrons_in_a_corbino_geometry,ballistic_to_hydrodynamic_transition_and_collective_modes_for_two_dimensional_electron} is still an open question in experiments. 
Although the generally admitted route $l_{ee}< W$, combined with $l_{ee} < l_{e}$, is the easiest situation where hydrodynamic flow can be decoupled from other effects, finding materials with high $l_{e}/l_{ee}$ ratios is a major issue~\cite{viscous_electron_fluids,hydrodynamic_approach_to_two_dimensional_electron_systems,phonon_mediated_hydrodynamic_transport_in_a_weyl_semimetal}. Thus, the relevance of the new alternative routes, compatible with viscous flow, greatly expands the possible scenarios where traditional hydrodynamic features may occur or even new unexpected phenomena may arise.


\begin{methods}
\label{methods}

The BTE is solved using a finite element method~\cite{a_finite_element_technique_for_solving_first_orderr_PDEs_in_Lp}. Unlike the nonlocal conductivity formalism~\cite{parahydrodynamics_from_weak_surface_scattering_in_ultraclean_thin_flakes,robustness_of_vorticity_in_electron_fluids}, it can be implemented for arbitrary boundary conditions. 
In a very long channel, the electric potential splits as $V(x,y) = V_H(x) + y \partial_y V$, where 
$V_H(x)$ is the Hall potential and $\partial_y V $ is a constant potential gradient. Therefore, $g = g(x,\theta)$ does not depend on $y$ and the BTE~\eqref{BTE} reduces to
\begin{align}
i\,\frac{g}{l_\omega} & + \cos \theta\, \partial_x \left( g- \frac{e V}{m {\rm v}_F} \right) + \sin \theta \,\partial_y V \nonumber \\
& + \frac{\partial_\theta g}{l_B} + \frac{g}{l_e} + \frac{g-g^{ee}}{l_{ee}} = 0 \ .
\label{BTEmethods}
\end{align}
We approximate the potential $V_H$ by its expansion in a truncated basis $\lbrace \phi_n (x)\rbrace_{n=1}^N$ of tent functions defined on $[-W/2,W/2]$ as $V_H(x) = \sum_{n=1}^N V_n \phi_n(x)$ and write the solution as  
\begin{equation}
    g(x,\theta) = \sum_{n=1}^N \sum_{m=1}^M g_{nm} \phi_n(x) \varphi_m (\theta) \ ,
\end{equation}
where $\lbrace \varphi_m (\theta)\rbrace_{m=1}^N$ is a periodic basis of tent functions defined on $[0,2\pi)$. We achieved convergence for $N \gtrsim 40$ and $M\gtrsim 32$. We found the weak form of Eq.~\eqref{BTE} and used a conforming Galerkin method to write the $N \times M$ system of linear equations. The resulting system is sparse in the spatial part and includes the boundary conditions for the scattered electrons. No-trespassing condition ($u_x=0$) is enforced, unless within a magnetic field, where the solution was later corrected to ensure no-trespassing condition. For 2D arbitrary geometries, the same procedure is applied to Eq.~\eqref{BTE}. We set the $g(\theta )$ distribution at the contacts, far away from the studied region, using as an input the result for simulations in uniform channels. We rewrite the equations at all other edges for the corresponding boundary conditions. We use $N \sim 10^4$ spatial elements for arbitrary geometries, with a triangle size $\leq 0.2 W$ and an adaptive method with a thinner meshing near the corners, and a smaller $M \sim 16$ to reduce computational cost. 

The NSE was solved analytically in a very long channel (see Section~\ref{analytical} of the SI) and numerically in other geometries using the finite element method~\cite{the_finite_element_method_for_elliptic_problems}, with triangular~\cite{locally_optimal_delaunay_refinement_and_optimisation_based_mesh_generation} Taylor-Hood elements \cite{space_time_taylor_hood_elements_for_incompressible_flows}. We impose mixed boundary conditions~\cite{boundary_conditions_of_viscous_electron_flow}, used the analytical solution in a channel to set the velocities at the contacts and imposed a constant potential and zero current flow at the metallic contacts. The fourth-order Runge-Kutta method was used to compute the streamlines. The NSE and DRE errors are defined comparing to the BTE results as $\varepsilon_{NS} \equiv 2 |I_\mathrm{NSE}-I_\mathrm{BTE}| / |I_\mathrm{NSE}+I_\mathrm{BTE}|$ and $\varepsilon_{DR} \equiv 2 |I_\mathrm{DRE}-I_\mathrm{BTE}| / |I_\mathrm{DRE}+I_\mathrm{BTE}|$ where $I = - e n \int_{-W/2}^{W/2} u_y {\, \rm d} x $ is the total current, proportional to the area under the velocity profile. In order to identify the diffusive regions we compute the quantity $1-\varepsilon_{DR}$ which indicates where the DRE equation is correct.

\end{methods}

\noindent{\bfseries References}\setlength{\parskip}{12pt}%

\bibliography{main}

\begin{addendum}

  \item [Acknowledgments] We wish to acknowledge R. Brito, G. Oleaga, A. Cortijo, J. Bernabeu, B. Zhou, E. Diez and M. Amado for discussions. This work was supported by the “(MAD2D-CM)-UCM” project funded by Comunidad de Madrid, by the Recovery, Transformation and Resilience Plan, and by NextGenerationEU from the European Union and Agencia Estatal de Investigaci\'{o}n of Spain (Grants PID2019-106820RB-C2 and PID2022-136285NB-C31). J.~E. acknowledges support from the Spanish Ministerio de Ciencia, Innovaci\'{o}n y Universidades (Grant FPU22/01039).

  \item[Competing Interests] The authors declare that they have no competing financial interests.
 
  \item[Correspondence] Correspondence and requests for materials should be addressed to jorgestr@ucm.es.

\item[Author contributions] J.E. did the analytical and numerical calculations. J.E., F.D-A and E.D. analyzed the results and wrote the manuscript.
F. D-A and E.D supervised the project. 

    \item[Data and code availbaility] The data and the computer code supporting the findings of this study are available from the corresponding author upon request.
 
\end{addendum}

\beginedfigures

\renewcommand{\theequation}{S.\arabic{equation}}

\setcounter{equation}{0}

\newpage

\noindent{\bfseries \LARGE Supplementary Information}\setlength{\parskip}{12pt}%

\section{Boltzman transport equation requirements}
\label{BTEvalidity}

In this work we assumed that the BTE is exact. As a consequence, we computed the NSE's and DRE's error by comparing their predictions against the BTE. Now, we discuss the requirements for the BTE to be valid. Let us start with Eq.~\eqref{BTEtimeDependent}, a semiclassical description that requires\cite{ashcroft_solid_state_physics} $e\mathcal{E} a \ll E_g^2 / E_F$, $\hbar \omega_c \ll E_g^2 / E_F$ and $\hbar \omega \ll E_g $, where $a$ is the lattice parameter, $\mathcal{E}$ is the typical electric field, $E_g$ is the energy gap and $E_F$ the Fermi energy. Using data for graphene~\cite{superballistic_flow_of_viscous_electron_fluid_through_graphene_constrictions} at $T= 100 \, \rm K$ and an electron density $n = 10^{12} \, \rm cm^{-2}$, we find $e \mathcal{E} a / E_F \sim 10^{-7} \ll 1 $ for an electric field $\mathcal{E} \sim 700 \, \rm V m^{-1}$ corresponding to a graphene ribbon of width $500 \, \rm nm$ carrying a current $I = 10 \, \rm \mu A$ in the diffusive regime. Also $\hbar \omega_c / E_F  \sim 0.05 \ll 1$ for a magnetic field $B = 1 \, \rm T$, and the last two conditions are met. The model also assumes non-quantitized energy levels, and excludes quantum phenomena such as the Shubnikov-de Haas effect~\cite{ashcroft_solid_state_physics,geometric_control_of_universal_hydodynamic_flow_in_a_two_dimensional_electron_fluid,boundary_scattering_in_ballistic_graphene} occurring at very low temperatures and high magnetic fields. Last, we implicitly assumed that we either have an ensemble of electrons or holes, but, if we were close to the neutrality point we could renormalize the model as in~\cite{boundary_conditions_of_viscous_electron_flow}.  Hence, we can safely assume Eq.~\eqref{BTEtimeDependent} for the purposes of this work. 

Let us now discuss Eq.~\eqref{BTE}. As explained in the main text, we assumed the molecular chaos approximation under which the scattering rates in $\Gamma \left[ \hat{f} \right]$ are a function of the $\hat{f}$ distribution at time $t$~\cite{di_ventra_electrical_transport_in_nanoscale_systems}. To proceed, we assumed Callaway's ansatz \cite{probing_carrier_interactions_using_electron_hydrodynamics,hyrdodynamic_electron_flow_in_high_mobility_wires} which is also known as a two relaxation time approximation, using two parameters $\tau_e$ and $\tau_{ee}$ to describe all collisions. More collisions operators are described in reference~\cite{probing_carrier_interactions_using_electron_hydrodynamics}, which may result in slight changes in the velocity profiles, but they do not drastically affect the electric current. Indeed, the global response mainly depends in the relaxation rates of the first two harmonics to the equilibrium distribution which are considered in Callaway's ansatz. We assumed $|\hat{g} | \ll {\rm v}_F$. For instance, in a $500\, \rm nm$-wide graphene channel under the previous conditions and carrying a high current $I = 10 \, \rm \mu A$, the drift velocity is $u_y \sim 10^4 {\rm \, m \, s^{-1}} \ll {\rm v}_F$. Also, thermal broadening of the Fermi distribution is small, and the equilibrium distribution even at $T \sim 300 \, \rm K$ is close to a step function. So we can assume that all interesting phenomena happens near the Fermi surface and $g$ is enough to characterize the electric response. We also assumed that the electron density is uniform. The second term in Eq.~\eqref{BTEtimeDependent} with $\nabla_{\mathbf{r}} f^{e}$ can be neglected in favour of $\nabla_{\mathbf{r}} V$ 
%
%
when $\pi {\rm v}_F \hbar \alpha /(2 e k_F) \ll 1 $, where $\alpha$ is the ratio between the density of carriers and the bias potential. For graphene under the previous conditions and $\alpha \sim 6 \times 10^{14} \, \rm V^{-1} m^{-2}$ as a typical value for a $300 \, \rm nm$-thick dielectric hBN layer \cite{dielectric_properties_of_hexagonal_boron_nitride_and_transition_metal_dichalcogenides_from_monolayer_to_bulk}, $\pi {\rm v}_F \hbar \alpha/(2 e k_F) = 0.003 \ll 1 $. If this condition were not met, the model would still be valid, but we would have to deal with electrochemical potentials in spite of having this straightforward derivation. Therefore, the use of Eq.~\eqref{BTE} is clearly justified in this work. 


\section{Tomographic regime \label{tomographicSection}}

In this paper, we mainly work under Callaway's ansatz~[23]
and assume that the relaxation time $\tau_{ee}$ and, consequently, the mean free path $l_{ee}$ is the same for all harmonic modes of the distribution function.
However, other works [30,31]
analyze the influence of long-lived odd moments of the distribution function on 
the ballistic-to-hydrodynamic crossover, the so called tomographic regime. They also show a fractional expansion similar to the one in the nonlocal conductivity formalism [26].
In this section we revisit this intermediate regime within our approach. Our formalism allows us to distinguish the relaxation rates for the terms with an even ($l_{ee}^{e}$) or an odd ($l_{ee}^{o}$) $m$ coefficient in a harmonic expansion of the form $g(\theta) = \sum_n c_n \cos (m \theta) + s_n \sin (m \theta)$. 

In this paper, we mainly work under Callaway's ansatz~\cite{Callaway1959} and assume that the relaxation time $\tau_{ee}$ and, consequently, the mean free path $l_{ee}$ is the same for all harmonic modes of the distribution function.
However, other works \cite{collective_modes_in_interacting_two_dimensional_tomographic_fermi_liquids, tomographic_dynamics_and_scale_sependent_viscosity_in_2D_electron_systems} analyze the influence of long-lived odd moments of the distribution function on 
the ballistic-to-hydrodynamic crossover, the so called tomographic regime. They also show a fractional expansion similar to the one in the nonlocal conductivity formalism~\cite{robustness_of_vorticity_in_electron_fluids}. 
In this section we revisit this intermediate regime within our approach. Our formalism allows us to distinguish the relaxation rates for the terms with an even ($l_{ee}^{e}$) or an odd ($l_{ee}^{o}$) $m$ coefficient in a harmonic expansion of the form $g(\theta) = \sum_n c_n \cos (m \theta) + s_n \sin (m \theta)$. 

Figure \ref{fig:tomographic}(a) and (b) show the distribution $g(\theta)$ in the tomographic regime and the velocity profiles for several values of $l_{ee}^{e}$ and $l_{ee}^{o}$. The tomographic regime emerges when $l_{ee}^{e} < W$ but $l_{ee}^{o} > W$, namely, only the even harmonics are allowed to relax. The velocity profile is curved in this regime, quite similar to a Poiseuille flow. However, odd terms in $g(\theta)$ do not relax, giving an excess of electrons travelling parallel to the channel, similarly to what occurs in the ballistic regime. Thus, the behaviour is not collective and as a result the NSE fails.
Figure \ref{fig:tomographic}(c) presents a color map for the NSE error respect the BTE exact solution. The region where DRE model is valid to account for the diffusive regime of transport is also shown.
Here, an extension of the validity of the NSE when $l_{ee}^{e}$ is combined with other effects, such as inelastic collisions $l_e$ or any relaxation mechanisms for the odd harmonics $l_{ee}^{o}$, is shown. Yet the consideration of $l_{ee}^{e}$ is not enough to enter the hydrodynamic by its own. 




\section{Derivation of the Navier-Stokes equation from the Boltzmann transport equation}
\label{NSEfromBTE}

Viscous electron flow is characterized by a collective behaviour which can be described with macroscopic variables such as $u_x, u_y, w_x$ and $w_y$ [see the introduction of equation~\eqref{armonics2}]. As a consequence, we can exactly expand the distribution $g(\theta)$ as a Fourier series and identify the first coefficients as the macroscopic variables 
\begin{equation}
\label{armonicExpan}
    g = u_x \cos \theta + u_y \sin \theta + w_x \cos 2 \theta + w_y \sin 2 \theta + \sum_{n=3}^\infty (c_n \cos n \theta + s_n \sin n \theta ) \ .
\end{equation}

After neglecting all terms that are not related to macroscopic variables, the harmonic expansion Eq.~\ref{armonics2} considered by the NSE is found. A similar series up to second order is introduced in Ref. \citenum{ballistic_and_hydrodynamic_magnetotransport_in_narrow_channels}. 
An alternative way to derive the NSE relies on the Chapman-Eskong expansion and computing several integrals in terms of the first two powers of the momentum. This is similar to consider the first coefficients of the previous expansion~(\ref{armonicExpan}). 

Here we show how to derive the NSE from the BTE by exploiting the advantages of keeping the electron dynamics close to the Fermi surface. This results in a simplification of the derivation of the NSE in comparison with conventional fluids. 
Let us first substitute the two harmonic approximation~\eqref{armonics2} in the BTE~\eqref{BTE}. After some straightforward algebra, we equal the coefficients for the first polar harmonics
\begin{subequations}
\begin{eqnarray}
\nabla \cdot {\bm u} &= 0 \ ,\label{comp_0}\\
\frac{1}{2}\left(  \begin{matrix}
\partial_x w_x + \partial_y w_y \\
\partial_x w_y - \partial_y w_x \end{matrix} \right)-\frac{e \nabla V}{m {\rm v}_F} + \frac{{\bm u} \times {\bm z}}{l_B} + \frac{{\bm u}}{\tilde{l}_e} &= 0 \ ,  \label{comp_u}\\
\frac{1}{2}\left(  \begin{matrix}
\partial_x u_x - \partial_y u_y \\
\partial_x u_y + \partial_y u_x \end{matrix} \right) + \frac{2 {\bm w}\times {\bm z}}{l_B} + \frac{{\bm w}}{\tilde{l}_{ee}} &= 0 \ ,\label{comp_w} 
\end{eqnarray}
\label{compNS}%
\end{subequations}
where $\tilde{l}_e = (l_e^{-1} + i l_\omega^{-1} )^{-1}$ and $\tilde{l}_{ee} = (l_e^{-1} + l_{ee}^{-1} +  i l_\omega^{-1} )^{-1} $ are the effective mean free paths. The NSE provides an approximate solution to Eqs.~\eqref{compNS} up to error terms in $\cos 3 \theta$ and $\sin 3 \theta$, but no more equations can be imposed. This can also be regarded as a finite element method in the angular basis $\lbrace \cos \theta , \sin \theta , \cos 2 \theta , \sin 2 \theta \rbrace$. We aim at eliminating $w_x$ and $w_y$ from the model, so we solve the linear system in Eq.~\eqref{comp_w} to find $w_x$ and $w_y$ in terms of the partial derivatives of the velocity field
\begin{align}
    w_x & = - 2\, \frac{\nu}{{\rm v}_F} \Big(  \partial_x u_{x}-\partial_y u_{y}  \Big) + 2\, \frac{\nu_H}{{\rm v}_F} \Big(\partial_x u_{y}+\partial_y u_{x} \Big)\ ,\\
    w_y & = - 2\, \frac{\nu}{{\rm v}_F} \Big(\partial_x u_{y}+\partial_y u_{x}   \Big) - 2\, \frac{\nu_H}{{\rm v}_F} \Big(\partial_x u_{x}-\partial_y u_{y} \Big)\ ,
\end{align}
where $\nu$ is the viscosity and $\nu_H$ is the Hall viscosity defined in Eq.~\eqref{viscosity} of the main text. Their dependence on the magnetic field is also a direct consequence of the BTE and does not require any additional reasoning as in Ref.~[\citenum{negative_magnetoresistance_in_viscous_flow_of_two_dimensional_electrons}]. We then write $\partial_x \omega_x$, $\partial_y \omega_x$, $\partial_x \omega_y$ and $\partial_y \omega_y $ and substitute in Eq.~\eqref{comp_u} to find the result Eq.~\eqref{NS}. 
Following this procedure we arrive at the best possible approximation to the BTE using two harmonics and it enables us to write the phenomenological parameters of the model in terms of microscopic ones, and to understand electron flow in terms of fundamental physical processes.

By comparing NSE and BTE results within the same physical scenarios one can demonstrate the validity of the NSE when any of the following conditions is fulfilled
\begin{eqnarray}
(l_{ee} < W) \text{ or } (l_{e} < W) \text{ or } (l_B < W) \text{ or } (l_\omega < W) \nonumber \\ \mathrm{\ \ or\ equivalently\ \ }  \nu <  {\rm v}_F W\,
\label{NSrequirement}
\end{eqnarray}
being the first one the widely accepted one as the hydrodynamic regime. Actually these conditions can be summarized in the general requirement $\nu <  {\rm v}_F W$ for the viscosity Eq.~(\ref{viscosity}) for most of the cases. If a magnetic field is present, the condition $\nu_H < {\rm v}_F W$ needs also to be fulfilled to avoid commensurability effects.
It is worth mentioning that the validity of the considered NSE does not directly imply a hydrodynamic regime of transport, due to the non-conventional dissipative term included in solid-state systems. Still after neglecting the region where the DRE is also valid, the NSE accuracy is a good fingerprint for viscous electron flow.

\section{Boundary conditions}
\label{BoundaryCondition}

For the sake of simplicity we will show the boundary conditions at the $x = W/2$ edge of the very long channel in figure~\ref{fig:pseudoFermi}. No-trespassing condition imposes a zero net flow of electrons through a boundary, namely $u_x = 0$, which is assumed in all models discussed in this work. Let us now comment separately the two boundary conditions considered in the main text.

\textit{Diffusive edge (DF)}. In this case, electrons are uniformly scattered in all directions  \cite{visualizing_poiseuille_flow_of_hydrodynamic_electrons,boundary_conditions_of_viscous_electron_flow}. This imposes the following condition for scattered electrons in the BTE equation
\begin{equation}
g(\theta) = 0\ , \qquad \pi/2 < \theta < 3\pi/2 \ .
\label{boundaryBTEdiffusive}
\end{equation}
However, the NSE is not based on the $g(\theta)$ distribution, so we need to rewrite this condition in terms of $u_x$ and $u_y$. In order to do that, firstly we must discuss some previous considerations. Let us assume that all electrons reaching the edge have suffered at least one inelastic collision inside the main flux in a previous time. Hence, its distribution is the same as in a Drude diffusive model, $g(\theta)\propto \sin\theta$. Notice that, within this assumption, we keep no trace from collisions at other edges. Additionally, the electrons that may have suffered a previous elastic collision have relaxed to a local equilibrium distribution with a drift velocity dependent on the distance to the edge. Therefore, those electrons contribute to the incident distribution with a term proportional to $\sin \theta$ and, after integrating over all distances, the expected incident distribution still fulfills such condition. Under all these considerations, we propose to use the following ansatz
\begin{equation}
g(\theta) = \hat{u}_y \sin \theta \ ,\qquad -\pi/2 < \theta < \pi/2\ .
\label{ansatzWall}
\end{equation}
Nevertheless, it is worth noticing that Eq.~\eqref{ansatzWall} may fail when more  complex processes are involved such as those in the case of the commensurability effect discussed in the main text.

By combining Eqs.~\eqref{boundaryBTEdiffusive} and~\eqref{ansatzWall} one can define the final distribution for all possible angles as follows
\begin{equation}
    g(\theta ) = \begin{cases} \hat{u}_y \sin \theta\ , & \text{ if } -\pi/2 \leq \theta \leq \pi / 2 \ ,\\
    0 \ ,& \text{otherwise} \ ,\end{cases} \ \ \Rightarrow \ \ g(\theta ) \simeq  \hat{u}_y\left(\frac{1}{2}\sin \theta +  \frac{4}{ 3\pi}  \sin 2 \theta \right)\ .
\label{finalG}   
\end{equation}
Last, since the NSE only admits two harmonics~\eqref{armonics2}, we rewrite Eq.~\eqref{finalG} by way of its Fourier series up to second order.  Here $\hat{u}_y$ is not the drift velocity but an effective velocity associated to incident electrons. The fact that the scattered electrons have lost their momenta yields $ u_y = \hat{u}_y/2$. Most importantly, by using Eq. \eqref{armonics2} we identify $u_y$ and $w_y$ and find the following ratio
\begin{equation}
    u_y = \frac{3\pi}{8} w_y\ . 
   \label{wyuy}
\end{equation}
However, $w_y$ is not a variable in the final model so we manage to eliminate it due to the channel translational symmetry that results in $\partial_y u_x = 0$ and $\partial_y u_y = 0$ and the continuity equation $ \partial_x u_x = 0$. Under these conditions, Eq.~\eqref{comp_w} can be solved for $w_y$ as follows 
\begin{equation}
w_y = -2 \nu  \partial_x u_y\ , 
 \label{compSolvedOut}
\end{equation}
where the viscosity $\nu$ is defined by Eq.~\eqref{viscosity}. Finally, by considering  Eqs.~\eqref{wyuy} and \eqref{compSolvedOut} we write the boundary condition only in terms of the model variables
\begin{equation}
\partial_x u_y = -\frac{u_y}{\xi}  \ \ \text{ where } \ \  \xi \equiv \frac{\nu}{{\rm v}_F} \frac{3\pi}{4} \ .
\label{boundaryNSdiffusive}
\end{equation}

Here we have defined the slip length $\xi$, which is usually assumed as a phenomenological parameter in the literature, but as we have demonstrated, it can be derived from more fundamental considerations. Most importantly, our formalism generalizes previous analysis\cite{boundary_conditions_of_viscous_electron_flow} by including the effect of inelastic collisions, magnetic field and ac driving. Remarkably, we have defined  local boundary conditions only in terms of quantities at the edge, and not at other inner positions of the channel as previously done\cite{ballistic_and_hydrodynamic_magnetotransport_in_narrow_channels}. This enables us to relate the boundary condition with the material parameters and the scattering mechanisms at the edge, and also eases generalization to other geometries. Note that a diffusive boundary does not yield a no-slip condition $\xi = 0 $ since incident electrons may have a drift velocity.

\textit{Partial specular edge (PS)}. The second kind of boundary condition corresponds to partial specularity (PS), where electrons reverse its normal momentum when they hit the boundary but keep part of its incident momentum, as if they partly reflected in a mirror. In order to model this PS boundary, we consider a previously proposed formalism~\cite{boundary_conditions_of_viscous_electron_flow}, express the magnitudes of interest in polar coordinates and proceed with the integration over $k$. Then, the equation for $g(\theta)$ in the BTE reads
\begin{equation}
g \left( \theta \right) = g \left( {\pi}-\theta \right) + d \,\cos \theta \left[  g \left({\pi}-\theta \right)   -\tfrac{2}{\pi}  \int_{\pi/2}^{3\pi/2} \cos^2 \theta' g\left({\pi}-\theta' \right) \, {\rm d} \theta' \right] 
\label{boundaryBTEspecular}
\end{equation}
for $\pi/2 < \theta < 3\pi/2$. Here $d \equiv \sqrt{\pi} h^2 h' k_F^3 \lesssim 1 $ is the dispersion coefficient defined in the text. The lower the dispersion coefficient $d$, the higher the specularity of the boundary. Once again we use the ansatz~\eqref{ansatzWall} for the incident distribution and Eq.~\eqref{boundaryBTEspecular} to find the distribution for all angles 
\begin{equation}
    g(\theta ) = \begin{cases} \hat{u}_y \sin \theta\ , & \text{ if } -\pi/2 \leq \theta \leq \pi / 2\ , \\
    \hat{u}_y \sin \theta + \hat{u}_y d \sin \theta \cos \theta \ , & \text{ otherwise} \ , \end{cases}  
\end{equation}
and consequently
\begin{equation}
   g(\theta) \simeq \hat{u}_y \left(1-\frac{2d}{3\pi} \right)  \sin \theta + \frac{\hat{u}_y d}{ 4} \sin 2 \theta \\ \Rightarrow \\ u_y = \left(\frac{4}{d} - \frac{8}{3\pi} \right) w_y
\end{equation}

By following the same procedure as for the DF edge, we reach a similar final condition but with a different value of the slip length as follows
\begin{equation}
\xi = \frac{\nu}{{\rm v}_F} \left( \frac{8}{d} - \frac{16}{3\pi} \right) \ .
\label{boundaryNSspecular}
\end{equation}
Note that in this case perfect-slip condition, $\xi \to \infty$, corresponds to $d=0$.  
\section{Polar plots of the $g(\theta)$ distribution for an extended set of parameters}
\label{PolarExtended}

Figure~\ref{fig:moreFermi} presents a collection of polar plots of the $g(\theta )$ distribution in different positions along a uniform channel of width $W$. The effect of the boundary conditions, the elastic and inelastic collisions, and the magnetic field on the detailed structure of $g(\theta )$ is considered. 
Viscous electron flow can be easily distinguished in these plots. Indeed, such collective behaviour is described using only macroscopic variables, and reveals itself by a regular distribution function $g(\theta)$, which can be approximated with two angular harmonics~\eqref{armonics2}. Conversely, the non-viscous behaviour presents sharper features that remain beyond the NSE model. Accordingly, in the latter physical scenario the quantitative comparison between the transport properties predicted by the NSE and the BTE clearly disagree.
\section{Velocity profile for an extended set of parameters} 
\label{ProfileExtended}

Figure~\ref{fig:moreProfiles} and~\ref{fig:crenellated} show a collection of transverse velocity profiles in uniform and crenellated channels under different physical conditions.

\section{Drude model}
\label{DRE_validity}
Let us derive the well-known Drude model as another approximation to the BTE~\cite{ashcroft_solid_state_physics} in which we further neglect the angular details of $g(\theta)$ and impose
\begin{equation} 
g(\theta) = u_x \cos \theta + u_y \sin \theta \ .
\label{armonics1}
\end{equation}
After substituting Eq.~\eqref{armonics1} in Eq.~\eqref{BTE} and equal the coefficient of the first harmonics, we find
\begin{subequations}
\begin{eqnarray}
 \nabla \cdot {\bm u} &=& 0 \ ,\label{DR_continuity}\\
 \omega_B \, {\bm u} \times {\bm z} + \frac{1}{\tilde{\tau}_e}\, {\bm u} &=& \dfrac{e }{m} \, \nabla V \ , \label{DR_momentum} 
\end{eqnarray}
\end{subequations}
where $1/\tilde{\tau}_e=\mathrm{v}_F/l_e+i\omega$. 

Approximation~\eqref{armonics1} is more restrictive than \eqref{armonics2} and thus, conditions for the validity of the DRE are clearly stricter than those of the NSE, namely
\begin{equation}
l_e \ll W \text{ or } l_\omega \ll W\ ,
\label{DRrequirement}
\end{equation}
while $l_B \ll W$ is not enough when $l_{ee} < W$.  
Remarkably in comparison to the NSE~\eqref{NS_momentum}, viscous terms proportional to $\nu$ and $\nu_H$ are absent in Eq.~\eqref{DR_momentum}, so this model cannot deal with elastic collisions effects. 
Indeed, even when elastic collisions are negligible, $l_{ee}\gg W$, the DRE fails when $l_e \sim W$ because it does not even account for collisions against the boundary. 

Figures ~\ref{fig:extendeMagnetoPlots} and~\ref{fig:cyclotronPlots} show the NSE and DRE error compared to the BTE in panels (a) and (b), respectively. As shown in all considered cases, the NSE description, where the viscosity accounts both for inelastic and elastic collisions, is much more reliable than that based on the DRE. 

Let us remark on the similarities of this result with phonon transport with phonons obeying a bosonic distribution function.
Despite this fact, hydrodynamic models have been shown to give precise results~\cite{derivation_of_a_hydrodynamic_heat_equation_from_the_phonon_boltzmann_equation_for_general_semiconductors,hydrodynamic_thermal_transport_in_silicon_at_temperatures_raging_from_100_to_300_K} when inelastic collisions are frequent, both because they cause a collective behaviour, but also because it improves the results of a purely diffusive model, analogously to what we discussed for electrons in this section. 
\section{Model error of the total current in a uniform channel for an extended set of parameters} 
\label{ErrorExtended}

Figure~\ref{fig:extendeMagnetoPlots} and~\ref{fig:cyclotronPlots} show a collection of color maps representing the NSE error respect to the exact BTE model in uniform channels of width $W$ under several physical conditions.


\section{Model parameters} \label{ModelParameters}

Let us discuss the values of the model parameters for some materials commonly used in experiments on electron hydrodynamics. We used this parameters in the discussion of the results. Table~\ref{tab:modelParametersGraphene} displays the elastic and inelastic mean free paths in hBN-encapsulated single layer graphene for several temperatures and electron density. We took the values from Ref.~[\citenum{superballistic_flow_of_viscous_electron_fluid_through_graphene_constrictions}], where the measured $l_{ee}$ is comparable with the theoretical prediction, and $l_e$ is of the order of magnitude of other experiments \cite{negative_local_resistance_caused_by_viscous_electron_backflow_in_graphene}. However, it strongly depends on the sample, and we wanted to estimate what happened with a less pure sample on a $\rm SiO_2$ substrate, so we reduced the zero-temperature mobility one order of magnitude~\cite{boron_nitride_substrates_for_high_quality_graphene_electronics}. We overprinted these parameters for several temperatures and electron density $n = 10^{12} \, \rm cm^{-2}$, on $200\, \rm nm $ (G$\mathrm{_{SiO_2}}$) and $500 \, \rm nm$-wide (G$\mathrm{_{hBN}}$) channels. We use the Fermi velocity ${\rm v}_F = 10^6 \, \rm m\,  s^{-1}$ to characterize the graphene band structure~\cite{the_electronic_properties_of_graphene}. 
Table \ref{tab:modelParametersGaAs} includes the values for a 2D electron gas in GaAs, estimated after the experimental results in Ref.~\cite{geometric_control_of_universal_hydodynamic_flow_in_a_two_dimensional_electron_fluid}. We display them for a $200 \, \rm n m$-wide channels at temperatures $T = 10$ and  $40 \, K$ and electron density $n = 2.45 \times 10^{11} \, \rm cm^{-2}$. The effective mass $m = 6.1 \times 10^{-32} \, \rm kg $ is used~\cite{geometric_control_of_universal_hydodynamic_flow_in_a_two_dimensional_electron_fluid}. 



\section{Deviation of the Hall field predicted by NSE vs BTE}
\label{HallVoltage}

As shown in previous works~\cite{visualizing_poiseuille_flow_of_hydrodynamic_electrons,ballistic_and_hydrodynamic_magnetotransport_in_narrow_channels}, the Hall field is also a significant physical magnitude to analyze the existence of hydrodynamic transport. Remarkably 
within our approach, based on the comparison between the NSE and the BTE results, the consideration of the current profiles, the Hall voltage, the total current, or the total Hall voltage yields the same conclusions, as proven in figures  \ref{fig:moreProfiles} and \ref{fig:errorPlotHall}.




\section{Analytical solution in a long channel}
\label{analytical}
The NSE and the DRE are amenable of an analytical solution in the limiting situation of a very long channel, shown in Fig.~\ref{fig:pseudoFermi}, where the electricpotential is $V(x,y) = V_H(x) + y\partial_y V(x,y) $, being $\partial_y V $ constant. Due to translational symmetry $\partial_y u_y = 0 $ and as a consequence of the continuity Eq.~\eqref{NScontinuity}
$\partial_x u_x = 0 $. Last, due to no-trespassing condition at the boundary, we can safely assume $u_x = 0$ all over the channel. 

First, we find the velocity profile $u_y(x)$ and the Hall potential $V_H(x)$ that obey the following simplified version of the NSE~\eqref{NS} 
\begin{subequations}
\begin{align}
  \omega_B u_y &= \frac{e}{m}\,\partial_x V\\
  -\nu \partial^2_x u_y + \frac{1}{\tilde{\tau}_e}\,u_y & = \frac{e}{m}\,\partial_y V\ .
\end{align}
\end{subequations}
These equations can be solved by taking into account any of the considered boundary conditions \eqref{boundaryNSdiffusive} or \eqref{boundaryNSspecular} in terms of the slip-length $\xi$ defined in Eq. \eqref{slipLength}. The solutions finally read as follows 
\begin{subequations}
\begin{eqnarray}
u_y = \frac{e \tilde{\tau}_e}{m} \left( 1 -\dfrac{\cosh  \dfrac{ x }{D_\nu} }{\cosh   \dfrac{ W }{2D_\nu}+ \dfrac{\xi}{D_\nu} \sinh   \dfrac{ W }{2D_\nu} } \right) \partial_y V \ ,\label{solutionNSuy}\\
V =  \left[y +\frac{e B \tilde{\tau}_e }{m}\left( x -\dfrac{ D_\nu  \sinh  \dfrac{ x }{D_\nu} }{\cosh    \dfrac{ W }{2D_\nu} + \dfrac{\xi}{D_\nu} \sinh   \dfrac{W}{2D_\nu} } \right)\right]\partial_y V\ ,
\end{eqnarray}
\label{NSsolChannel}%
\end{subequations}
where $D_\nu \equiv \sqrt{\nu \tilde{\tau}_e}$. The total Hall voltage and the current along the channel can be obtained as $V_H = - I \, B /(e n)$ and $I = (L/R)\partial_y V$, respectively, where the resistance is
\begin{equation}
R =   \frac{mL}{e^2 n\tilde{\tau}_e}\, \,  \left( W -\frac{2 D_\nu }{\coth    \dfrac{ W }{2D_\nu}+ \dfrac{\xi}{D_\nu} } \right)^{-1}\ .
\end{equation}
The Drude model can be solved in a similar way such that the DRE in a very long channel reads
\begin{subequations}
\begin{align}
\omega_B u_y &= \frac{e}{m}\,\partial_x V \ ,\\
 \frac{u_y}{\tilde{\tau}_e} & = \frac{e}{m}\,\partial_y V\ .
\end{align}
\end{subequations}
Once again, no-trespassing boundary condition is the only meaningful one. Indeed the solution in a very long channel is the same as in an infinite two dimensional plane, since edge scattering does not play any role in this model  
\begin{subequations}
\begin{align}
u_y &= \frac{e\tilde{\tau}_e}{m}\,\partial_y V \ ,\\
V &=  \left(y+  \frac{eB \tilde{\tau}_e}{m} \, x\right) \partial_y V\ .
\end{align}
\end{subequations}
The expressions in the DRE for $I$ and $V_H$ hold with the resistance being
\begin{equation}
R = \frac{m}{e^2 \tilde{\tau}_e n} \frac{ L}{W}\ .
\end{equation}


\section{Hydrodynamic features in different flow geometries}
\label{Geometry}

In this section we present NSE simulations in nonuniform geometries where some relevant and experimentally proven hydrodynamic features arise.
Figure~\ref{fig:superballistic} shows evidences of the so-called superballistic electron flow in point contacts and crenellated channels. In all cases the electricresistance of the 2D devices is clearly diminished when the new hydrodynamic conditions(~\ref{NSrequirement}) are fulfilled.
Figure~\ref{fig:whirlpool} represents the electronic flow in a Hall bar device, where the existence of whirlpools due to the hydrodynamic transport results in negative resistance. Contrary to DRE that cannot predict any whirlpool or negative resistance~\cite{visualizing_poiseuille_flow_of_hydrodynamic_electrons,negative_local_resistance_caused_by_viscous_electron_backflow_in_graphene}, we demonstrate that NSE perfectly reproduces these features when $l_{ee} \to \infty$ and inelastic collisions are the only relevant scattering events, as seen in figure~\ref{fig:whirlpool}. 



\begin{figure}[t]
\centerline{\includegraphics[width=17cm]{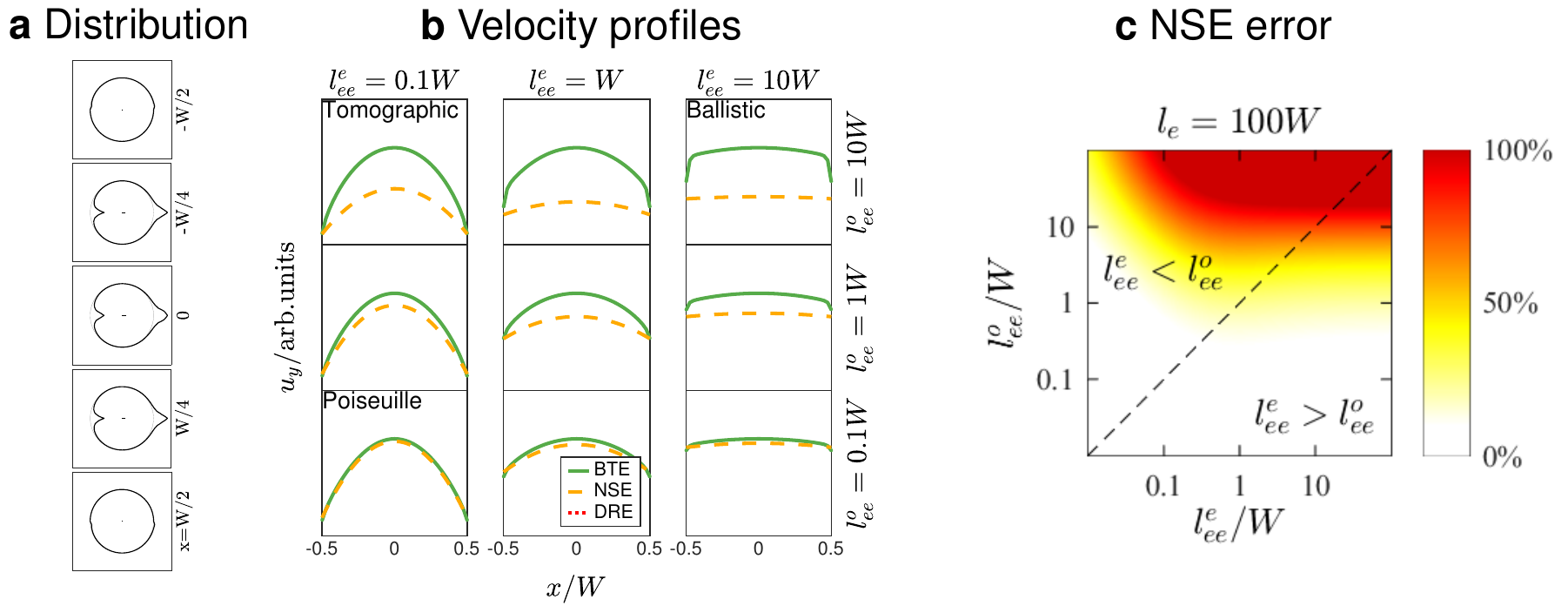}}
\caption{\bodyfigurelabel{fig:tomographic} Tomographic regime. (a) Fermi distribution in the tomographic regime where there are elastic collisions that relax the even terms in the distribution function, while $l_{ee}^{e} = 0.1 W$, $l_{ee}^{o} = 10W$ and $l_{e} = 100 W$ are negligible. The accumulation of electrons travelling parallel to the channel is a feature in common with the ballistic regime, which is due to the lack of relaxation mechanisms for the odd harmonics in the distribution function. A similar behaviour for the collective modes for a uniform channel of width $W$ and DF edges was observed in Ref.~[31].
(b) Velocity profiles when $l_e = 100W$ as a function of the even $l_{ee}^e$ and odd $l_{ee}^{o}$ mean free paths. (c) NSE error for the total current. This evidence demonstrates that  the tomographic regime cannot be described as a collective electron flow.
}
\label{moreprofiles}
\end{figure}

\begin{figure}[t]
\centerline{\includegraphics[width= 10cm]{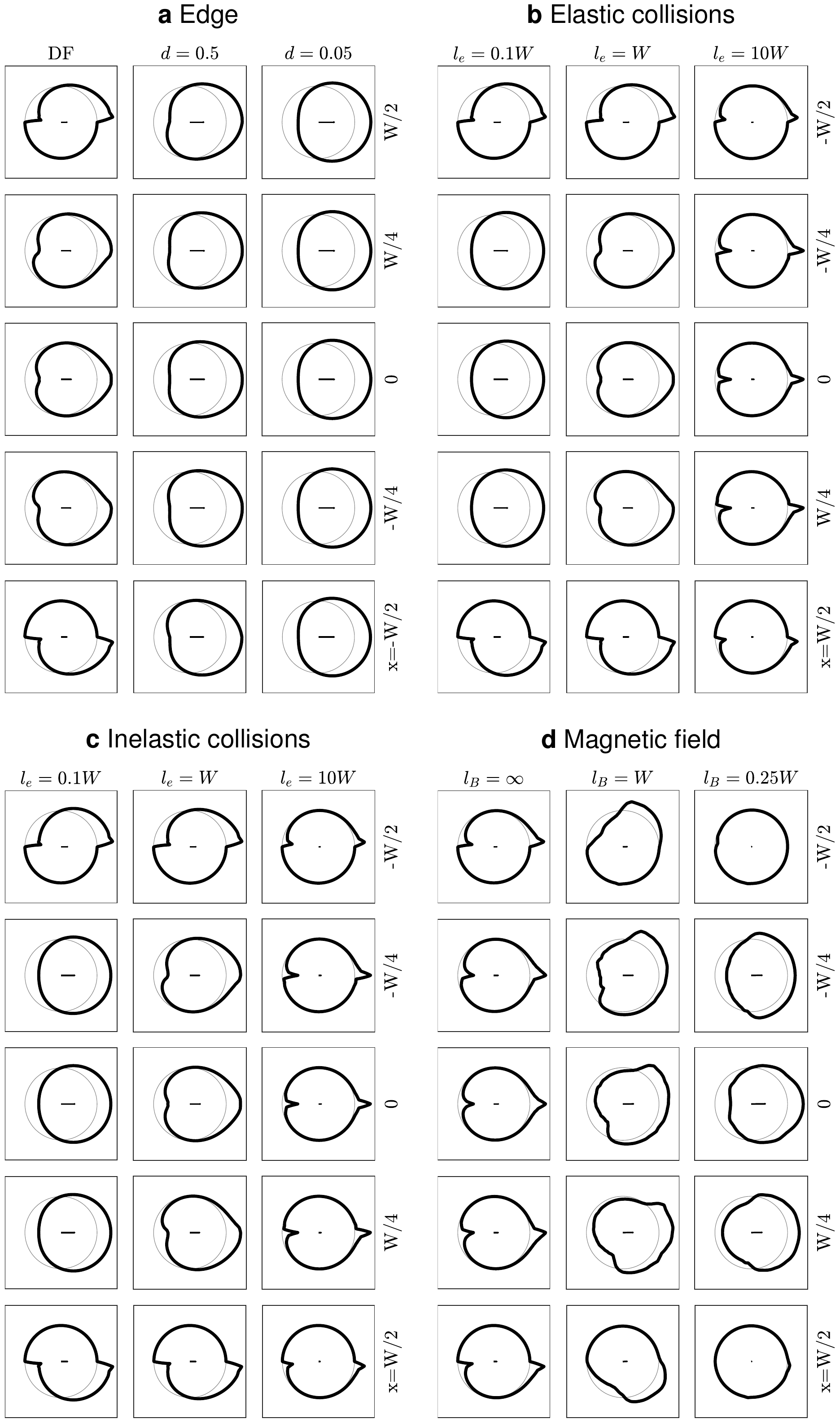}}
\caption{\bodyfigurelabel{fig:moreFermi} Polar plots of the distribution $g(\theta)$ as a function of the transverse position $x$ in a channel of width $W$. Different effects are remarked in each panel. 
(a)~Intermediate regime ($l_e=l_{ee}=W$) where magnetic effects are irrelevant for various boundary conditions, DF and PS edges with $d=0.5$ and $0.05$ as dispersion coefficients. Notice that the DF edge has the most uneven shape near the edges.
(b)~The relevance of elastic collisions is analyzed in terms of $l_e$ when $l_{ee} = 10 W$ in absence of field effects.
(c)~The relevance of inelastic collisions is analyzed in terms of $l_{ee}$ when $l_{e} = 10 W$ in absence of field effects.
(d)~The relevance of the magnetic field is analyzed in terms of $l_{B}$ when $l_{e} =l_{ee}= 10 W$ in the absence of an ac field.} 
\end{figure}

\begin{figure}[t]
\centerline{\includegraphics[width= 17cm]{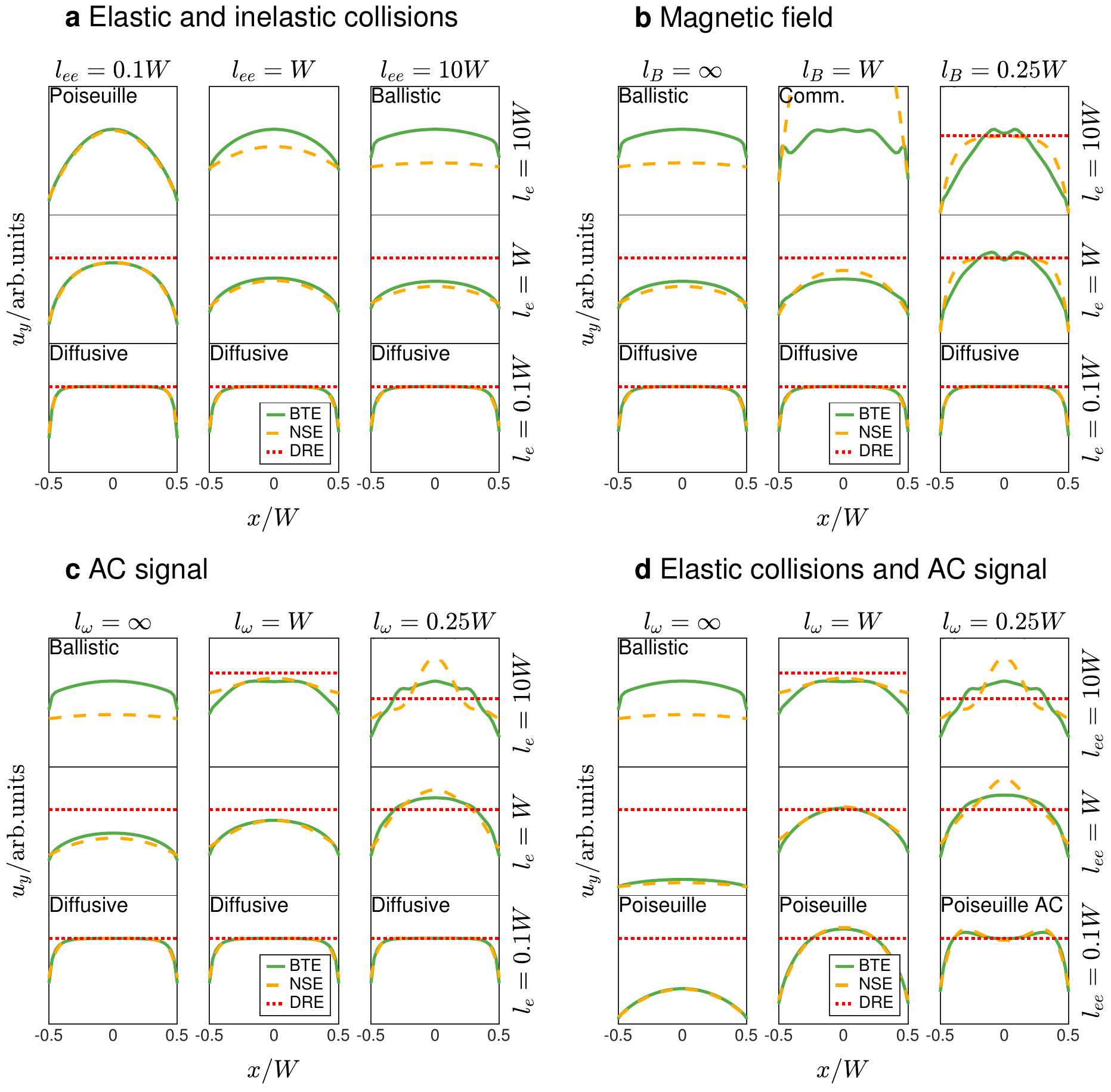}}
\caption{\bodyfigurelabel{fig:moreProfiles} Transverse velocity profiles in a  channel of width $W$ given by different transport regimes. Panel (a)~compares the velocity profiles for different values of the elastic $l_{ee}$ and inelastic $l_e$ mean free paths. Poiseuille, ballistic and diffusive regimes are discussed in the main text. Panel (b) compares the results for different magnitudes of the magnetic field. Due to the commensurability effect, the condition $l_B\simeq W$ results in an enhanced resistance and a larger error of the NSE and the DRE approaches. However, high magnetic fields $l_B = 0.25 W$ result in a curved profile and a NSE improved accuracy. 
Panel (c) and (d) compares the results for different magnitudes of the ac field and several magnitudes of  $l_{e}$ and $l_{ee}$, respectively. Notice the occurrence of curved profiles when $l_\omega = 0.25 W$.}
\label{moreprofiles2}
\end{figure}

\begin{figure}[t]
\centerline{\includegraphics[width= 16cm]{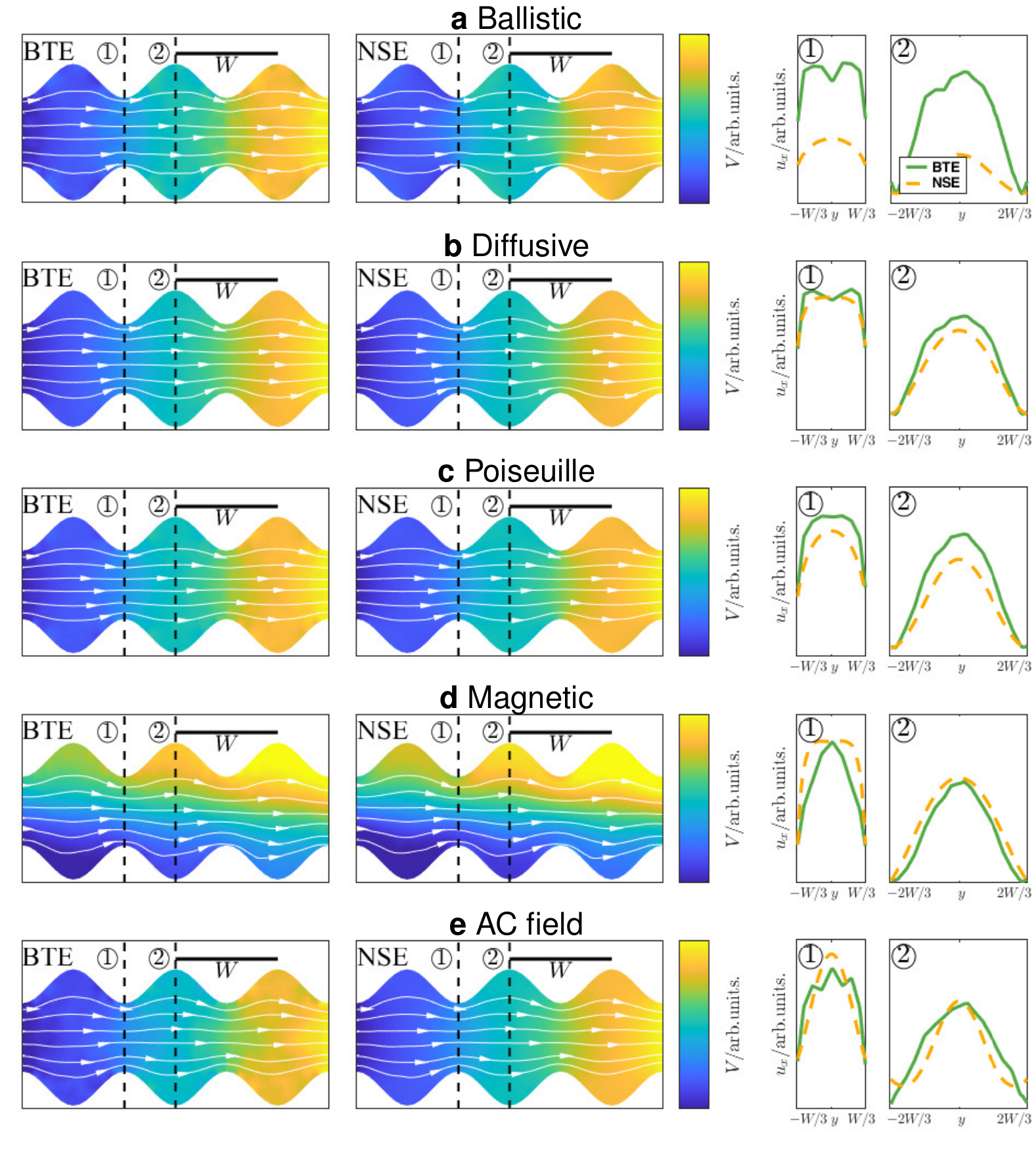}}
\caption{\bodyfigurelabel{fig:crenellated} Electron flow in a very long crenellated channel whose width oscillates between $2W/3$ and $4W/3$ with a spatial period $W$. We solved the BTE and the NSE for DF edges 
to compare the velocity profiles across two transverse sections, labelled \textcircled{1} and \textcircled{2} in the plots. (a) The hydrodynamic NSE does not match the exact BTE in the ballistic regime, $l_e = 2W$ and $l_{ee} = 2W$. These default values were used in the other panels if they are not explicitly changed. The NSE gives correct results under (b) frequent inelastic collisions $l_e = 0.25 W$, (c) elastic collisions $l_{ee} = 0.25 W$, (d) a magnetic field such that $l_B = 0.25W$ or (e) an AC field $l_\omega = 0.25W$.}
\end{figure}

\begin{figure}[t]
\centerline{\includegraphics[width= 17cm]{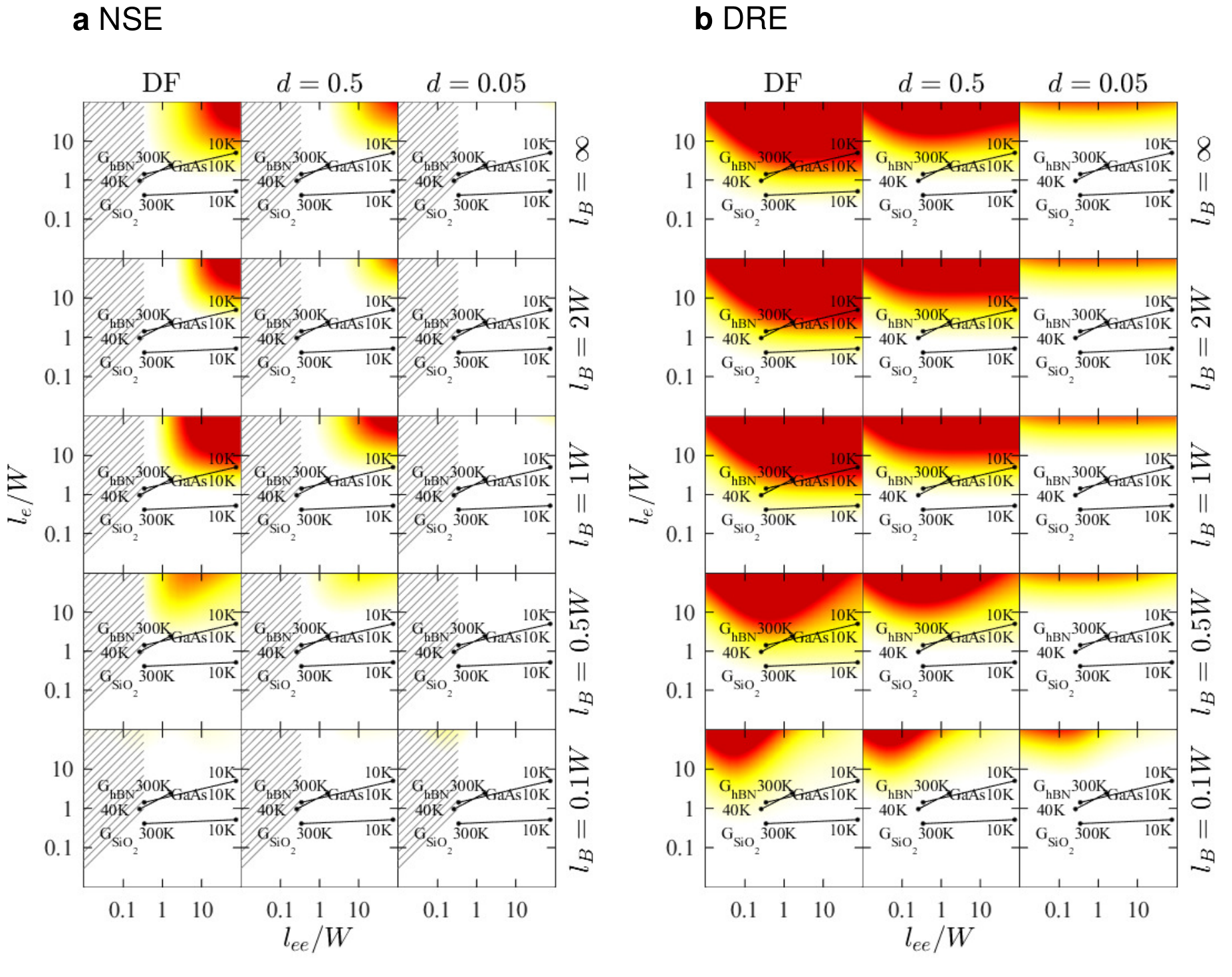}}
\caption{\bodyfigurelabel{fig:extendeMagnetoPlots} 
Color maps that represent the error of the total current modeled by (a)~the NSE and (b)~the DRE in comparison with the BTE results. Same colorbar as in Fig. \ref{fig:cyclotronPlots}, such that white regions are those where models can be safely applied. Maps are evaluated as a function of $l_e$ and $l_{ee}$ for different conditions in absence of an ac field. Each column accounts for a different type of edge, either DF or PS with a particular dispersion coefficient $d$. Each row accounts for a different magnetic field. Panel~(a) shows that, even at low magnitudes, the magnetic field improves the NSE accuracy, despite the exceptional case of the commensurability effect.
Moreover, a lower $d$, which correspond to more specular edges, increase the region of validity of the model. Panel~(b) shows the DRE accuracy. The commensurability effect is hidden by the larger error of this model. Increasing the magnetic field does enlarge the region of validity, although the model is still not valid for all values of $l_{ee}$.}
\end{figure}

\begin{figure}[t]
\centerline{\includegraphics[width= 17cm]{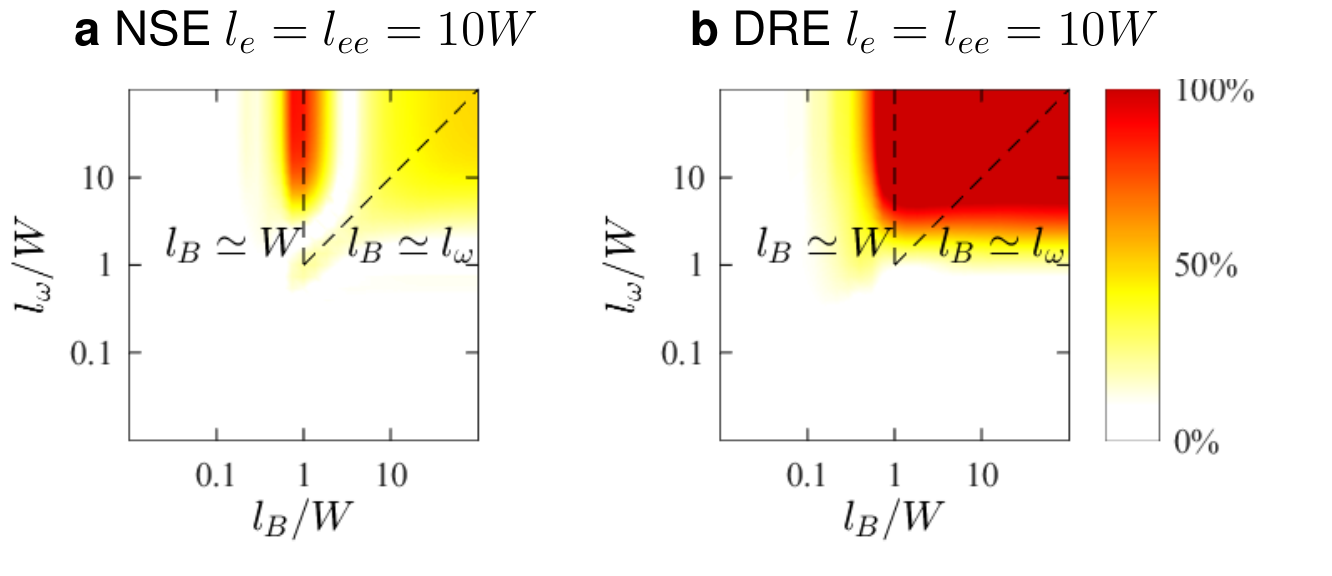}}
\caption{\bodyfigurelabel{fig:cyclotronPlots} 
Color maps of the NSE and the DRE versus $l_B$ and $l_\omega$. The evaluated magnitude is the total current in a uniform channel with DF edges. (a)~The NSE error when inelastic and elastic collisions are more frequent than in Fig. (3b). Increasing the collision rates results in a decrease of the error regardless of $l_e$ and $l_{ee}$.
(b)~The DRE error under the same conditions of (a). The larger red area includes the commensurability and resonance effects, which are not particularly revealed in this model. The condition $l_\omega \ll W $ is still enough for the model to work, but the condition $l_B \ll W$ does not suffice for low $l_{ee}$ values, as seen in figure~\ref{fig:extendeMagnetoPlots}.}
\end{figure}

\begin{figure}[t]
\centerline{\includegraphics[width= 17cm]{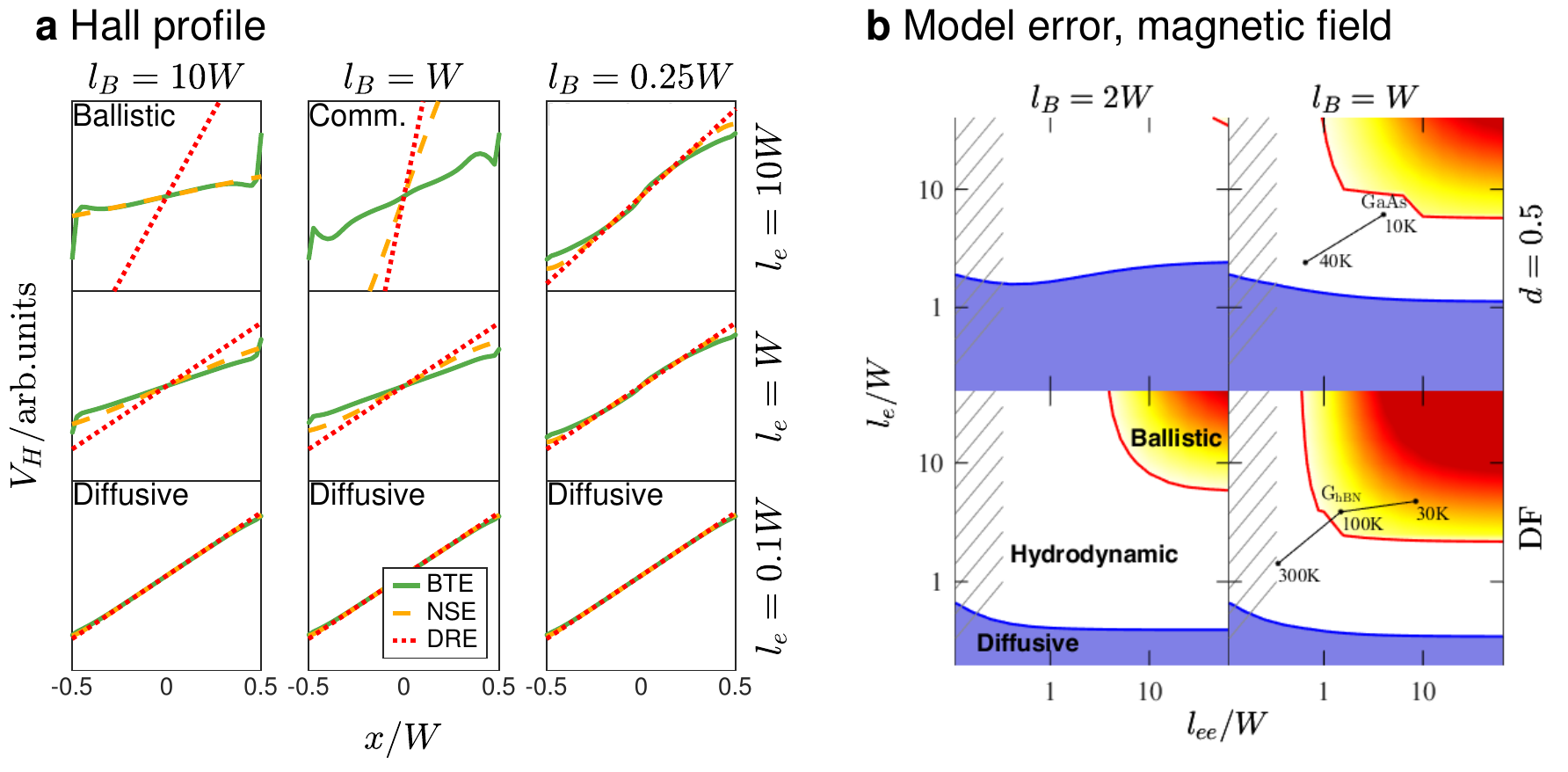}}
\caption{\bodyfigurelabel{fig:errorPlotHall} (a)~Hall potential $V_H$ for different magnitudes of
the magnetic field in a channel of width $W$ with DF edges. Panel (b) shows color maps that represent the NSE model error of the Hall voltage in comparison with the BTE results versus $l_e$ and $l_{ee}$ mean free paths, in absence of an ac field. Each plot
accounts for a DF or PS ($d = 0.5$) edge, and for a different cyclotron radius $l_B =\infty, 2W$ and
$W$. Grey patterned areas 
represent the conventional requirements
($l_{ee} \ll W$ and $l_{ee} \ll l_e$). Overprinted lines show typical physical conditions considered in
some experiments in a $W = 500 nm$ channel (see Section~\ref{ModelParameters} of the SI).
}
\end{figure}

\begin{figure}[t]
\centerline{\includegraphics[width= 17cm]{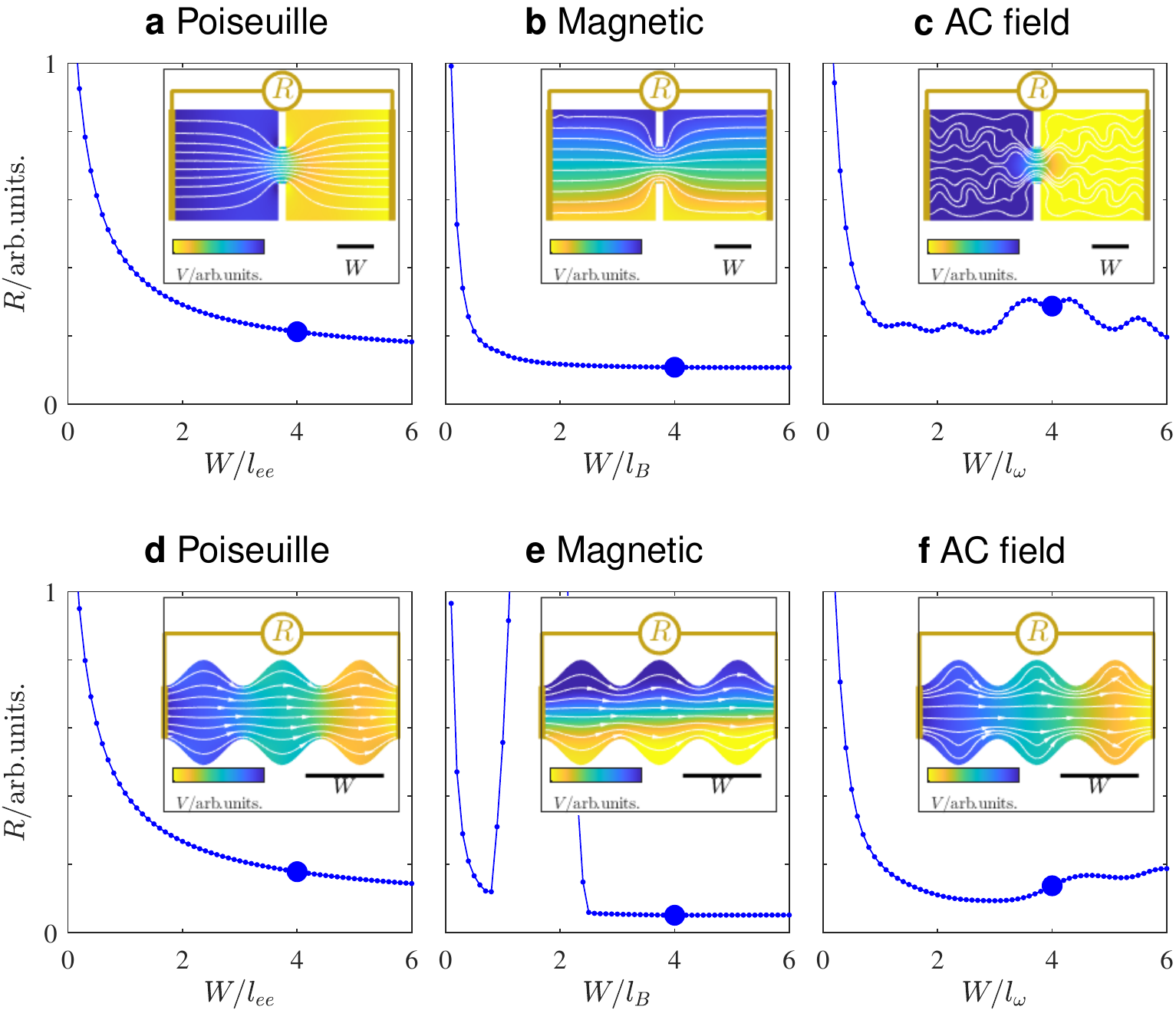}}
\caption{\bodyfigurelabel{fig:superballistic} NSE simulations of the superballistic electron flow through a point contact (a-c) and a crenellated channel (d-f). All panels assume negligible inelastic scattering $l_e = 10W$ and a DF edge. The electrical resistance and the NSE flow distribution for $W/l_{ee}=4$, marked with solid circle in each panel, are shown. (a,d) The increase of elastic collision reduces the electrical resistance when $l_{ee} < W$. (b,e) The increase of the magnetic field, in the absence of elastic collisions, also reduces the resistance, after the region $l_B \sim W$ where commensurability effects occur. (c,f)~In the absence of elastic collisions, an ac field such that $l_\omega < W$ reduces the real part of the impedance, associated with dissipation. The proposed three routes leads to a sizable reduction of the electric resistance below the ballistic limit.}
\end{figure}

\begin{figure}[t]
\centerline{\includegraphics[width= 17cm]{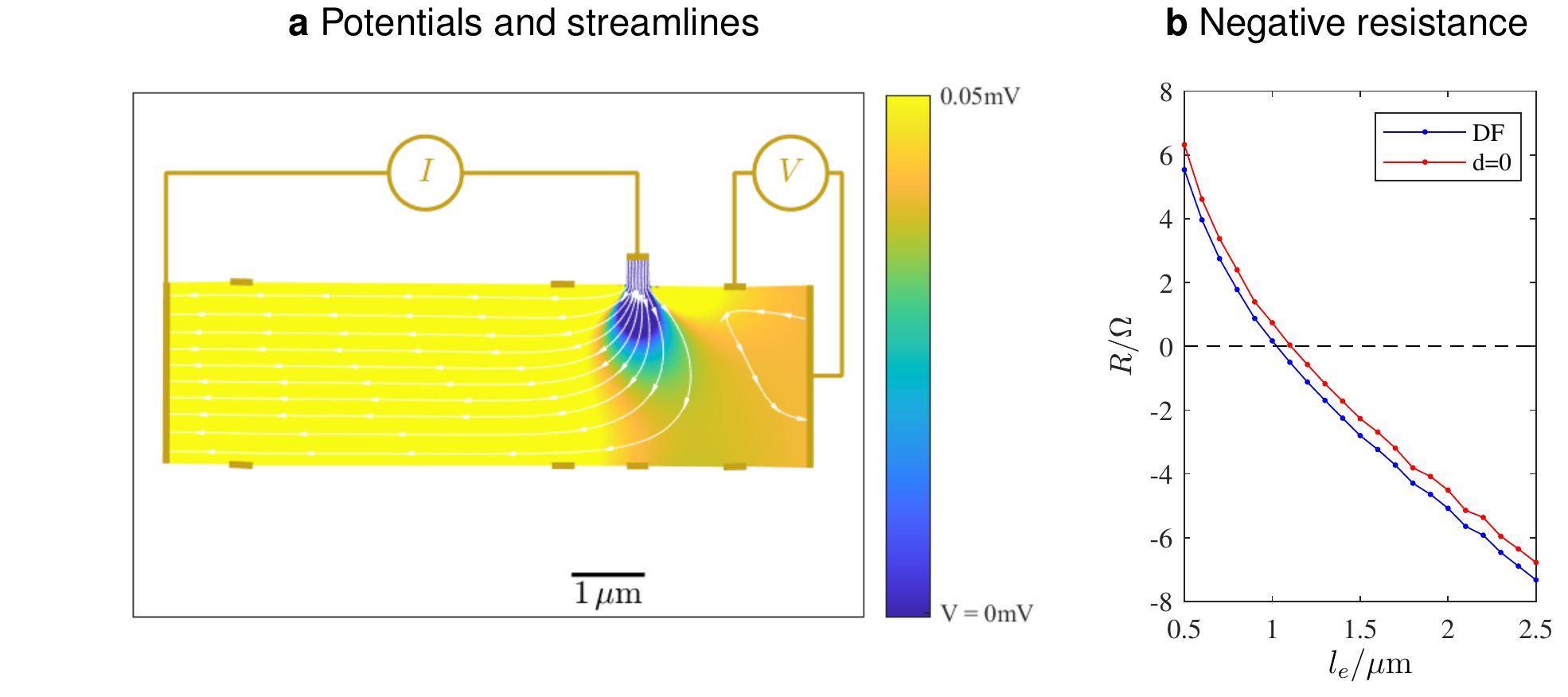}}
\caption{\bodyfigurelabel{fig:whirlpool} NSE simulation in a Hall bar device where the hydrodynamic behaviour results in negative resistance~[8].
We assume DF edges and the experimental geometry~[27], 
being $W=2.5 \, \rm \mu m$ the width of the bar and $L = 1 \, \mu m $ the separation between the injector and the first voltage probe. Elastic collisions at very low temperatures can be neglected $l_{ee} = \infty$. Inelastic collisions $l_e < \infty$ remain and its contribution to the viscosity $\nu$ results in whirlpools and negative resistance. (a) Potentials and streamlines of the viscous electron flow when $l_e = 2 \, \rm \mu m$, the total passing current is $I = 1 \, \rm \mu A$ and the negative resistance is $R = V/I = -6 \, \Omega $.
 (b) Negative resistance $R$ as a function of $l_e$. 
The NSE is valid due to the first alternative route, and predicts whirlpools if $l_e \sim 1 \, \rm \mu m $. 
However, the NSE does not predict the ballistic transition to positive values of $R$ established for values of $l_e$ larger than those shown in the figure.
 }
\end{figure}






\newpage

\begin{table}[t]
\caption{\label{tab:lengthScales}%
Table 1: Length scales in the BTE \eqref{BTE}. }
\begin{tabular}{cll}
Length &Expression &
\multicolumn{1}{c}{\textrm{Name}}\\
\hline
$l_e$ & ${\rm v}_F \tau_e$ & Inelastic mean free path \\
$l_{ee}$ & ${\rm v}_F \tau_{ee}$ & Elastic mean free path  \\
$l_B$  & $ m {\rm v}_F /eB = {\rm v}_F /\omega_B $ & Cyclotron radius   \\
$l_\omega$ & ${\rm v}_F / \omega$  & ac length  
\end{tabular}
\end{table}

\begin{table}[t]
\caption{\label{tab:modelParametersGraphene}%
Table 2: Model parameters for graphene approximated after the experimental results in Ref.~\citenum{higher_than_ballistic_conduction_of_viscous_electron_flows}. We show the inelastic $l_e$ and elastic $l_{ee}$ mean free paths for different temperatures $T$ and electron densities $n$. For the graphene on a $\rm SiO_2$ substrate we assume that the inelastic mean free path at $T=0 \, \rm K$ is one order of magnitude lower \cite{boron_nitride_substrates_for_high_quality_graphene_electronics} and use the same values for $l_{ee}$. }
\begin{tabular}{l|cccc|cccc} 
& \multicolumn{4}{c|}{\textrm{Graphene on hBN}} & \multicolumn{4}{c}{\textrm{Graphene on $\rm Si O_2$}}\\
& \multicolumn{2}{c}{$n = 10^{12} \, \rm cm^{-2}$} & \multicolumn{2}{c|}{$n = 2\times 10^{12} \, \rm cm^{-2}$}& \multicolumn{2}{c}{$n =  10^{12} \, \rm cm^{-2}$} & \multicolumn{2}{c}{$n = 2\times 10^{12} \, \rm cm^{-2}$}\\
$T/K $& $l_e/ \rm \mu m$ & $l_{ee} / \rm \mu m $ & $l_e/ \rm \mu m$ & $l_{ee} / \rm \mu m $ & $l_e/ \rm \mu m$ & $l_{ee} / \rm \mu m $ & $l_e/ \rm \mu m$ & $l_{ee} / \rm \mu m $ 
\\
\hline
10 & 2.6 & 38 &2.3 & 54     & 0.26& 38& 0.24 & 54\\
30 & 2.4 & 4.2 & 2.2 & 6.0    & 0.25 & 4.2& 0.24 & 6.0\\
100 & 1.9 & 0.74 & 1.7 & 1.0  & 0.25 &0.74&0.23 & 1.0\\
300 & 0.7 & 0.17 & 0.70 & 0.24  & 0.20 & 0.17 & 0.19 & 0.24\\
\end{tabular}
\end{table}

\begin{table}[t]
\caption{\label{tab:modelParametersGaAs}%
Table 3: Model parameters for a 2D electron gas in GaAs after the experimental measurements reported in Ref.~\citenum{geometric_control_of_universal_hydodynamic_flow_in_a_two_dimensional_electron_fluid}.  We show the inelastic $l_e$ and elastic $l_{ee}$ mean free paths for different temperatures $T$ and electron density $n$. } 
\begin{tabular}{l|cccc} 
& \multicolumn{4}{c}{\textrm{2D electron gas in GaAs}} \\
& \multicolumn{2}{c}{$n = 2.45 \times 10^{11} \, \rm cm^{-2}$} & \multicolumn{2}{c}{$n = 1.45 \times 10^{11} \, \rm cm^{-2}$} \\
$T/K $& $l_e/ \rm \mu m$ & $l_{ee} / \rm \mu m $ & $l_e/ \rm \mu m$ & $l_{ee} / \rm \mu m $
\\
\hline
10 & 1.22 & 0.80 & 0.65&0.37\\
20 & 0.90 & 0.25 & 0.43 & 0.17\\
40 & 0.48 & 0.13 & 0.30 & 0.12\\
\end{tabular}
\end{table} 

\end{document}